\documentclass{article}
\usepackage{amsmath, amssymb}
\usepackage{graphicx}
\usepackage{color}
\usepackage{textcomp}

\usepackage{graphics}

\renewcommand{\abstractname}{Abstract.}
\renewcommand\abstract{\hfil\break\topsep=0pt\partopsep=0pt\parsep=0pt\itemsep=0pt\relax
\trivlist\item[\hskip\labelsep
{\bfseries\abstractname}]\if!\abstractname!\hskip-\labelsep\fi}

\newcommand{\email}[1]{{(e-mail: #1)}}

\def\keywordname{{\bfseries Key words:}}
\def\keywords#1{\par\addvspace\baselineskip\noindent\keywordname\enspace
\ignorespaces#1}

\def\jelclassname{{{\bfseries JEL Classification:}} }
\def\jelclass#1{\par\addvspace\medskipamount\noindent\jelclassname\
\ignorespaces#1}

\def\subclassname{{\bfseries Mathematics Subject Classification (1991):} }
\def\subclass#1{\par\addvspace\medskipamount\noindent\subclassname\
\ignorespaces#1}

\def\title#1{\hfil\break\hfil\break
\hfil\break\par\addvspace\baselineskip\noindent
\ignorespaces{\LARGE\bf#1}\hfil\break}

\def\author#1{\par\addvspace\baselineskip\noindent
\ignorespaces{\large\bf#1}}

\def\institute#1{\par\addvspace\baselineskip\noindent
\ignorespaces{\small#1}\hfil\break}

 \newtheorem{thm}{Theorem}[section]

 \newtheorem{prop}[thm]{Proposition}
 \newtheorem{rem}[thm]{Remark}
 \numberwithin{equation}{section}
\newtheorem{exa}[thm]{Example}

\newcommand{\bm}{\bibitem}

\newcommand{\be}{\begin{equation}}
\newcommand{\ee}{\end{equation}}

\newcommand{\bea}{\begin{eqnarray}}
\newcommand{\eea}{\end{eqnarray}}
\newcommand{\beaa}{\begin{eqnarray*}}
\newcommand{\eeaa}{\end{eqnarray*}}

\newcommand{\UU}{{\mathcal U}}
\newcommand{\ZZ}{{\mathbb Z}}
\newcommand{\HH}{{\mathbb H}}

\newcommand{\RR}{{\mathbb R}}
\newcommand{\CC}{{\mathbb C}}

\begin{document}

\title{A General Asymptotic Implied Volatility \\
for Stochastic Volatility Models}

\author{Pierre Henry-Labord\`ere}

\institute{Barclays Capital
\\ \email{pierre.henry-labordere@barcap.com}}


\begin{abstract}
In this paper, we derive a general asymptotic implied volatility at
the first-order for any stochastic volatility model using the heat
kernel expansion on a Riemann manifold endowed with an Abelian
connection. This formula is particularly useful for the calibration
procedure. As an application, we obtain an asymptotic smile for a
{\it SABR model with a mean-reversion term}, called $\lambda$-SABR,
corresponding in our geometric framework to the Poincar\'e
hyperbolic plane. When the $\lambda$-SABR model degenerates into the
$SABR$-model, we show that our asymptotic implied volatility is a
better approximation than the classical Hagan-al expression
\cite{sab}. Furthermore, in order to show the strength of this
geometric framework, we give an exact solution of the SABR model
with $\beta=0$ or $1$. In a next paper, we will show how our method
can be applied in other contexts such as the derivation of an
asymptotic implied volatility for a  Libor  market model with a
stochastic volatility (\cite{phl2}).

\end{abstract}


\jelclass{G13}

\subclass{58J65}


\section{Introduction}

Since 1973, the Black-Scholes formula \cite{bla}  has been
extensively used by traders in financial markets to price options.
However, the original Black-Scholes derivation is based on several
unrealistic assumptions which are not satisfied under real market
conditions. For example, in the original Black-Scholes framework,
assets are assumed to follow log-normal processes (i.e. with a
constant volatility). This hypothesis can be relaxed by introducing
more elaborate models called local and stochastic volatility models
(see \cite{gat} for a nice review).

\vskip 3truemm

\noindent On the one hand, local volatility models assume that the
volatility $\sigma_L(f,t)$ depends only on the underlying $f$ and on
the time $t$. The market is still complete and, as shown by Dupire
\cite{dup}, there is a unique diffusion term $\sigma_L(f,t)$ which
can be calibrated to the current market of European option prices.
On the other hand, stochastic volatility models assume that the
volatility itself follows a stochastic process \cite{hul}: in this
case, the market becomes incomplete as it is not possible to hedge
and trade the volatility with a single underlying asset. It can be
shown that the local volatility function represents some kind of
average over all possible instantaneous volatilities in a stochastic
volatility model \cite{gat}.

\vskip 3truemm

 \noindent For these two types of models (local and
stochastic), the resulting Black-Scholes partial differential
equation becomes complicated and only a few exact solutions are
known \cite{car,hes}. The most commonly used solutions are the
Constant Elasticity of Variance model (CEV) which assumes that the
local volatility function is given by $\sigma_L(f,t)=\sigma_0
f^\beta$ with $\sigma_0$ and $\beta$ constant, and the Heston  model
which assumes a mean-reverting square-root process for the variance.

\vskip 3truemm

\noindent In all other cases, analytical solutions are not available
and singular perturbation techniques have been used to obtain
asymptotic expressions for the price of European-style options (see
\cite{fou} for a review of singular perturbation techniques). As for
fast reverting volatility processes, similar expansions also exist
\cite{fou}.

\vskip 3truemm

 \noindent These singular perturbation techniques can
also be used to derive an implied volatility (i.e. {\it smile}). By
definition, this implied volatility is the value of the volatility
that when put in the Black-Scholes formula, reproduces the market
price for a European call option. In \cite{sab}, the authors
discover that the local volatility models predict the wrong behavior
for the smile: when the price of the underlying decreases
(increases), local volatility models predict that the smile shifts
to higher (lower) prices. This problem can be eliminated with the
stochastic volatility models such as the SABR model (depending on
$4$ parameters: Sigma, Alpha, Beta, Rho) \cite{sab}. The SABR model
 has recently been the focus of much attention as it provides a simple asymptotic
smile for European call options, assuming a small volatility.

\vskip 3truemm

\subsection*{{\it Outline}}

\vskip 3truemm

\noindent In this paper, we obtain asymptotic solutions for the
conditional probability and the implied volatility up to the
first-order with any kind of stochastic volatility models using the
heat kernel expansion on a Riemannian manifold. This asymp\-to\-tic
smile is very useful for calibration purposes. The smile at
zero-order (no dependence on the expiry date) is connected to the
geodesic distance on a Riemann surface. This relation between the
smile at zero-order and the geodesic distance  has already been
obtained in \cite{ber,ber1} and used in \cite{ave} to compute an
asymptotic smile for an equity basket. Starting from this nice
geometric result, we show how the first-order correction (linear in
the expiry date) depends on an Abelian connection which is a
non-trivial function of the drift processes.

\vskip 3truemm

\noindent We derive the  unified asymptotic implied volatility in
two steps:

\vskip 3truemm

\noindent First, we  compute the local volatility function
associated to our general stochastic volatility model. It
corresponds to the mean value of the stochastic volatility. This
expression depends on the conditional probability which satisfies by
definition a backward Kolmogorov equation. Rewriting this equation
in a covariant way (i.e. independent of a system of coordinates), we
 find a general asymptotic solution in the short-time limit using
the heat kernel expansion on a Riemannian manifold. Then, an
asymptotic local volatility at first-order is obtained using a
saddle-point method. In this geometric framework, the stochastic
(local) volatility model will correspond to the geometry of (real)
complex curves (i.e. Riemann surfaces). In particular, the SABR
model can be associated to the geometry of the (hyperbolic)
Poincar\'e plane $\HH^2$. This connection between $\HH^2$ and the
SABR model has been obtained in an unpublished paper \cite{len} and
presented in \cite{len1}.  Similar results can be found in
\cite{ber1}.

\vskip 3truemm

\noindent The second step consists in using a one-to-one asymptotic
correspondence between a local volatility function and the implied
volatility. This relation is derived using the heat kernel on a
time-dependent real line.

\vskip 3truemm

\noindent Next, we focus on a specific example and derive an
asymptotic implied volatility at the first-order for a SABR model
with a mean-reversion term which we call $\lambda$-SABR. The
computation for the smile at the zero-order is already presented in
\cite{ber1} and a similar computation for the implied volatility at
the first-order (without a mean-reversion term) is done in
\cite{sab1}.

\vskip 3truemm

\noindent Furthermore, in order to show the strength of this
geometric approach, we  obtain {\it two exact solutions} for the
conditional probability in the SABR model (with $\beta=0 \; ,1$).
The $\beta=0$ solution  has already been obtained in an unpublished
paper \cite{len} and rederived by the author in \cite{phl}. For
$\beta=1$, an extra dimension  appears and in this case, the SABR
model  is connected to the three-dimensional hyperbolic space
$\HH^3$. This extra dimension appears as a (hidden) Kaluza-Klein
dimension.

\vskip 3truemm

\noindent As a final comment for the reader not familiar with
differential geometry, we have included a short appendix explaining
some key notions such as manifold, metric, line bundle and Abelian
connection (see \cite{egu} for a quick introduction). The
saddle-point method is also described in the appendix.

\section{Asymptotic Volatility for  Stochastic Volatility Models}

A stochastic volatility model depends on two SDEs, one for the asset
$f$ and one for the volatility $a$. Let denote
$x=(x_i)_{i=1,2}=(f,a)$, with initial conditions
$\alpha=(\alpha_i)_{i=1,2}=(f_0,\alpha)$. These variables $x_i$
satisfy the following stochastic differential equations (SDE)

\bea &&dx^i =b_i(x,t)dt+\sigma_i(x,t) dW_i \label{se}\\
 &&dW_idW_j=\rho_{ij}(x,t)dt \eea

 \noindent with the initial condition $x_i(t=0)=\alpha_i$ (with $b^f=0$ in the risk-neutral measure
 as $f$ is a traded asset).

\noindent We have that the local volatility is the mean value of the
stochastic volatility \cite{gat}

\bea \sigma^2(f,\tau)=  { \int_{0}^\infty \sigma_1(\tau,x)^2
p(x,\tau|\alpha) {dx_2} \over \int_{0}^\infty p(x,\tau|\alpha)
{dx_2}} \label{mean1} \eea

\noindent with $p(x,\tau|\alpha)$ the conditional probability.

\noindent In order to obtain an asymptotic expression for the local
volatility function, we will use an asymptotic expansion
 for the conditional probability $p(x,\tau|\alpha)$ satisfying the
Backward Kolmogorov equation

\bea {\partial p \over \partial \tau}=b^i{\partial_i
p}+g^{ij}{\partial_{ij} p } \; , \;  (i,j)={a,f}\label{fp1} \eea
\bea p(\tau=0)=\delta(x-\alpha) \eea

\noindent with $g^{ij} \equiv {1 \over 2}\rho_{ij}\sigma_i\sigma_j
$. \noindent In (\ref{fp1}), we have used the Einstein summation
convention, meaning that two identical indices are summed. For
example, $\partial_i b^i p= \sum_{i=1}^2 \partial_i b^i p$. Note
that in the relation $g^{ij} \equiv {1 \over
2}\rho_{ij}\sigma_i\sigma_j $ although two indices are repeated,
there are no implicit summation over $i$ and $j$. As a result,
$g^{ij}$ is a symmetric tensor precisely dependent  on these two
indices. We will adopt this Einstein convention throughout this
paper.

\noindent In the next section, we will show how to derive  an
asymptotic conditional probability for any multi-dimensional
stochastic volatility models (\ref{se}) using  the heat kernel
expansion on a Riemannian manifold (we will assume here that $n$ is
not necessary equal to $2$ as it is the case for a stochastic
volatility model.). In particular, we will explain the DeWitt's
theorem which gives the asymptotic solution to the heat kernel. An
extension to the time-dependent heat kernel will be also given as
this solution is particulary important in finance to include term
structure.

\section{Heat Kernel Expansion}

\subsection{Heat kernel on a Riemannian manifold} In this section,
the partial differential equation (PDE) (\ref{fp1}) will be
interpreted as the heat kernel on a general smooth
\underline{$n$-dimensional manifold} $M$ (here we have that $i,j=1
\cdots n$) without a boundary, endowed with the \underline{metric}
$g_{ij}$ (see Appendix A for definitions and explanations of the
underlined names). The inverse of the metric $g^{ij}$ is defined by
\bea g^{ij}={1 \over 2}\rho_{ij} \sigma_i \sigma_j
\label{inversemetric}\eea \noindent and the metric ($\rho^{ij}$
inverse of $\rho_{ij}$, i.e. $\rho^{ij}\rho_{jk}=\delta^i_k$) \bea
g_{ij}=2 {\rho^{ij} \over \sigma_i \sigma_j} \label{metric} \eea
\noindent The differential operator \bea D=b^i\partial_i +
g^{ij}\partial_{ij} \label{d2}\eea which appears in (\ref{fp1}) is a
second-order elliptic operator of Laplace type. We can then show
that there is a unique \underline{connection} $\nabla$ on ${\cal
L}$, a \underline{line bundle} over $M$, and a unique smooth
function $Q$ on $M$ such that \bea D&=&g^{ij} \nabla_i \nabla_j+Q
\\&=&g^{-{1 \over 2}}(\partial_i +{\cal A}_i) g^{1 \over
2}g^{ij}(\partial_j +{\cal A}_j) +Q \label{d1}\eea

\noindent Using this connection, (\ref{fp1}) can be written in the
covariant way, i.e.

\bea {\partial  \over \partial \tau}p(x,y,\tau)=D p(x,y,\tau)
\label{hkk}\eea

\noindent If we take ${\cal A}_i=0 \;, Q=0$ then $D$ becomes the
Laplace-Beltrami operator (or Laplacian) $\Delta=g^{-{1 \over
2}}\partial_ig^{1 \over 2}g^{ij}\partial_j$. For this configuration,
(\ref{hkk}) will be called the Laplacian heat kernel.

\noindent We may express the connection ${\cal A}^i$ and $Q$ as a
function of the drift $b_i$ and the metric $g_{ij}$ by identifying
in (\ref{d1}) the terms $\partial_i$ and $\partial_{ij}$ with those
in (\ref{fp1}). We find

\bea {\cal A}^i&=&{1 \over 2}(b^i-g^{-{1 \over
2}}\partial_j(g^{1/2}g^{ij})) \label{aaa} \\
Q&=&g^{ij}({\cal A}_i{\cal A}_j-b_j {\cal A}_i-\partial_j {\cal
A}_i) \label{qqq}\eea

\noindent Note that the Latin indices $i$,$j \cdots$ can be lowered
or raised using the metric $g_{ij}$ or its inverse $g^{ij}$. For
example ${\cal A}_i=g_{ij} {\cal A}^j$.

\noindent The components ${\cal A}_i=g_{ij} {\cal A}^j$ define a
local one-form ${\cal A}={\cal A}_i dx^i$. We deduce that under a
change of coordinates $x^{i'}(x^j)$, ${\cal A}_i$ undergoes the
vector transformation  $ {\cal A}_{i'} \partial_i x^{i'} = {\cal
A}_{i}$. Note that the components $b_i$ don't transform as a vector.
This results from the fact that the SDE (\ref{se}) has been derived
using the Ito calculus and not the Stratanovich one.

\noindent Next, let's introduce the Christoffel's symbol
$\Gamma_{ij}^k$ which depends on the metric and its first
derivatives

\bea \Gamma_{ij}^k={1 \over 2} g^{kp} (\partial_j g_{ip}+\partial_i
g_{jp}-\partial_pg_{ji}) \label{chr} \eea

\noindent (\ref{aaa}) can be re-written

\bea {\cal A}^i&=&{1 \over 2}(b^i-g^{pq}\Gamma_{pq}^i)  \eea

\noindent Note that if we define  \bea
p'=e^{\chi(x,\tau)-\chi(x=\alpha,\tau=0)}p \label{gt} \eea then $p'$
satisfies the same equation as $p$ (\ref{hkk}) but with \bea {\cal
A}'_i &\equiv& {\cal
A}_i-\partial_i \chi \label{gc}\\
{Q}' &\equiv& Q+\partial_\tau \chi \eea

\noindent The transformation (\ref{gt}) is called a {\it gauge
transformation}. The reader should be aware that the transformation
(\ref{gc}) only applies to the connection ${\cal A}_i$ with lower
indices.

\noindent The constant phase $e^{\chi(x=\alpha,\tau=0)}$ has been
added in (\ref{gt}) so that $p$ and $p'$ satisfy the same boundary
condition at $\tau=0$. Mathematically, (\ref{gt}) means that $p$ is
a \underline{section} of the line bundle $\cal L$ and when we apply
a (local) Abelian gauge transformation, this induces an action on
the connection $\cal A$ (\ref{gc}) (see Appendix A). In particular,
if the one-form $\cal A$ is exact, meaning that there exists a
smooth function $\chi$ such that ${\cal A}=d\chi$ then the new
connection ${\cal A}'$ (\ref{gc}) vanishes.

\vskip 3truemm

\noindent The asymptotic resolution of the heat kernel (\ref{hkk})
in the short time is an important problem in theoretical physics
 and in mathematics. In physics, it corresponds to the solution of the
Euclidean Schrodinger equation on a fixed space-time background
\cite{dewitt} and in mathematics, the heat kernel - corresponding to
the determination of the spectrum of the Laplacian - can give
topological information (e.g. the Atiyah-Singer's index theorem)
\cite{gil}. The following theorem proved by
Minakshisundaram-Pleijel-De Witt-Gilkey gives the complete
asymptotic solution for the heat kernel on a Riemannian manifold.

\begin{thm}
Let M be a Riemannian manifold without a boundary. Then for each $x
\in M$, there is a complete asymptotic expansion

\bea p(x,y,\tau)={\sqrt{g(x)} \over (4\pi \tau)^{n \over
2}}\sqrt{\Delta(x,y)}{\cal P}(x,y)e^{-{\sigma(x,y) \over
2\tau}}\sum_{n=1}^\infty a_n(x,y)\tau^n \;, \; \tau \rightarrow 0
\label{shk}\eea
\end{thm}

\begin{itemize}

\item \noindent Here, $\sigma(x,y)$ is the Synge world function
equal to one half of the square of geodesic distance $|x-y|_g$
between $x$ and $y$ for the metric $g$. This distance is defined as
the minimiser of

\bea |x-y|^2_g=min_{C} \int_0^T g_{ij} {dx^i \over dt}  {dx^j \over
dt} dt \eea and $t$ parameterises the curve $C$. The Euler-Lagrange
equation gives the following geodesic differential equation which
depends on the Christoffel's coefficients $\Gamma_{j k}^i$
(\ref{chr})

\noindent \bea {d^2 x^i \over d^2 t} +\Gamma_{j k}^i {d x^j \over d
t} {d x^k \over d t}=0 \label{geodesics}\eea

\item \noindent $\Delta(x,y)$ is the so-called Van Vleck-Morette
determinant \bea \Delta(x,y)=g(x)^{-{1 \over 2}}det (- {\partial^2
\sigma(x,y) \over
\partial x
\partial y}  )g(y)^{-{1 \over 2}} \label{mor}\eea

\noindent with $g(x)=det g_{ij}(x,x)$

\item ${\cal P}(x,y)$ is the parallel
transport of the Abelian connection along the geodesic from the
point $y$ to $x$

\bea {\cal P}(x,y)=e^{\int_{C(x,y)} A_i dx^i} \label{pargauge} \eea

\item \noindent The $a_i(x,y)$ ($a_0(x,y)=1$) are smooth functions
on $M$ and depend on geometric invariants such as the scalar
curvature $R$. The first coefficients are fairly easy to compute by
hand. Recently $a_n(x,x)$ has been computed up to order $n=8$. These
formulas become exponentially more complicated as $n$ increases. For
example, the formula for $a_6$ has $46$ terms. The first
coefficients are given below

\bea
a_1(x,x)&=&P \equiv {1 \over 6} R+Q  \\
a_2(x,x)&=& {1 \over 180}(R_{ijkl} R^{ijkl}-R_{ij}R^{ij}+\Delta
R^2)\\&+&{1 \over 2} P^2+{1 \over 12} {\cal R}_{ij} {\cal R}^{ij}+{1
\over 6} \Delta P \eea

\noindent with $R_{ijkl}$ the Riemann tensor, $R_{ij}$ the Ricci
tensor and  $R$ the scalar curvature given by \bea
R^{i}_{jkl}&=&\partial_{l}\Gamma_{jk}^i-\partial_{k}\Gamma_{jl}^i+\Gamma^p_{jk}\Gamma_{pl}^i-\Gamma^p_{jl}\Gamma_{pk}^i
\\
R_{jl}&=&R^{i}_{jil}\\
R&=&g^{ij}R_{ij} \\ {\cal R}_{ij}&=&[\nabla_i,\nabla_j]\eea

\end{itemize}

\begin{rem}
In financial mathematics, when $p$ is computed at n$^{\mathrm th}$
order, we call the solution an asymptotic solution of order ${n
\over 2}$ as the asymptotic order is defined according to the
dimensionless parameter $(\sigma^2_0 t)^n$. For example,  the SABR
model (that we will review) is an asymptotic solution at the
second-order and  only  requires the knowledge of the first heat
kernel coefficient $a_1$.
\end{rem}

\noindent Let's now explain how to use the heat kernel expansion
(\ref{shk}) with a simple example, namely a log-normal process.

\begin{exa}[Log-normal process]
The SDE is $df=\sigma_0 f dW$ and using the definition
(\ref{metric}), we obtain the following one-dimensional metric

\bea g_{ff}={2 \over {\sigma^2_0 f^2}} \eea

\noindent When written with the coordinate $s=\sqrt{2}{ln(f) \over
\sigma_0}$ , the metric is flat $g_{ss}=1$ and all the heat kernel
coefficients depending on the Riemann tensor vanish. The geodesic
distance between two points $s$ and $s'$ is given by the classical
Euclidean distance $ d(s,s')=|s-s'| $ and the Synge function is
$\sigma(s,s')={1 \over 2}(s-s')^2$.

\noindent In the old coordinate $[f]$, $\sigma$ is \bea
\sigma(f,f_0)={1 \over \sigma^2_0}ln( {f \over f_0} )^2 \eea

\noindent Furthermore, the connection $\cal A$ (\ref{aaa}) and the
function $Q$ (\ref{qqq}) are given by

\bea &&{\cal A}=-{1 \over 2 f}df \\
&&Q=-{\sigma_0^2 \over 8} \eea

\noindent Therefore the parallel transport ${\cal P}(f,f_0)$ is
given by ${\cal P}(f,f_0)= e^{- {1 \over 2} ln({f \over f_0}) }$.

\noindent Plugging these expressions into (\ref{shk}), we obtain the
following first-order asymptotic solution for the log-normal process
\bea p_1(f,\tau|f_0)={1 \over {f} (2\pi \sigma_0^2 \tau)^{1 \over
2}}e^{-{ln( {f \over f_0})^2 \over 2 \sigma_0^2 \tau}- {1 \over 2}
ln({f \over f_0})}(1-{\sigma_0^2 \over 8}\tau+\cdots)
\label{hkln1}\eea

\noindent At the second-order, the second heat kernel coefficient is
given by $a_2={1 \over 2}Q^2={\sigma_0^4 \tau^2 \over 128}$ and the
asymptotic solution is

\bea p_2(f,\tau|f_0)={1 \over {f} (2\pi \sigma_0^2 \tau)^{1 \over
2}}e^{-{ln( {f \over f_0})^2 \over 2 \sigma_0^2 \tau}- {1 \over 2}
ln({f \over f_0})}(1-{\sigma_0^2 \over 8}\tau+{\sigma_0^4 \tau^2
\over 128}) \label{hkln2}\eea

\noindent Furthermore, we note that $\cal A$ and $Q$ are exact,
meaning that ${\cal A}=d \Lambda$, $Q=-\partial_{\tau} \Lambda$,
with $\Lambda=-{1 \over 2}ln(f)+{\sigma_0^2 \tau \over 8}$. Modulo a
gauge transformation $p'=e^{{1 \over 2}ln({f \over f_0})+{\sigma_0^2
\tau \over 8}}p$, $p'$ satisfies the (Laplacian) heat kernel on
$\RR$

\bea \partial_s^2 p'=\partial_{\tau} p \eea

\noindent whose solution is the normal distribution. So the exact
solution for $p=p'e^{-{1 \over 2}ln({f \over f_0})-{\sigma_0^2 \tau
\over 8}}$ is given by

\bea p(f,\tau|f_0)={1 \over {f} (2\pi \sigma_0^2 \tau)^{1 \over
2}}e^{-{(ln( {f \over f_0})+{\sigma^2_0 \tau \over 2} )^2 \over 2
\sigma_0^2 \tau}} \label{hkln}\eea
\end{exa}

\vskip 3truemm

\noindent To be thorough, we conclude this section with a brief
overview of the derivation of the heat kernel expansion.

\noindent We start with the Schwinger-DeWitt antsaz

\bea p(t,x|y)={\sqrt{g(x)} \over (4\pi \tau)^{n \over
2}}det[\Delta(x,y)]^{{1 \over 2}}e^{-{\sigma(x,y) \over 2\tau}}
{\cal P}(x,y) \Omega(t,x|y) \label{antsaz} \eea

\noindent Plugging (\ref{antsaz}) into the heat kernel equation
(\ref{hkk}), we derive a PDE satisfied by the function
$\Omega(t,x|y)$

\bea \partial_t \Omega=(-{1 \over t}\sigma^i \nabla_i+{\cal P}^{-1}
\Delta^{-{1 \over 2}} ({D}+Q) \Delta^{1 \over 2} {\cal P})\Omega
\label{eqomega} \eea

\noindent with $\nabla_i=\partial_i+{\cal A}_i$ and $\sigma_i=
\nabla_i \sigma$, $\sigma^i=g^{ij} \sigma_j$. The regular boundary
condition is $\Omega(t,x|y)=1$. We solve this equation by writing
the function $\Omega$ as a formal series in $t$:

\bea \Omega(t,x|y)=\sum_{n=0}^\infty a_n(x|y) t^n \label{series}\eea

\noindent Plugging this series (\ref{series}) into (\ref{eqomega})
and identifying the coefficients in $t^n$, we obtain an infinite
system of ordinary coupled differential equations:

\bea &&a_0=1 \\
&&(1+{1 \over k} \sigma^i \nabla_i) a_k = {\cal P}^{-1} \Delta^{-{1
\over 2}} ({D}+Q)\Delta^{1 \over 2} {\cal P} a_{k-1} \eea

\noindent The calculation of the heat kernel coefficients in the
general case of arbitrary background offers a complex technical
problem. The Schwinger-DeWitt's method is quite simple but it is not
effective at higher orders. By means of it only the two lowest order
terms were calculated. For other advanced methods see \cite{avr}.

\subsection{Heat kernel on a time-dependent Riemannian manifold}

In most financial models, the volatility diffusion-drift terms can
explicitly depend  on time. In this case, we obtain a time-dependent
metric and connection. It is therefore useful to generalise the heat
kernel expansion from the previous section to the case of a
time-dependent metric defined by

\bea g_{ij}=2{\rho^{ij}(t) \over \sigma_i(t) \sigma_j(t)} \eea

\noindent This is the purpose of this section.

\noindent The differential operator \bea {\cal D}=b^i(t)\partial_i +
g^{ij}(t)\partial_{ij} \label{d2}\eea which appears in (\ref{fp1})
is a time dependent family of operators of Laplace type. Let $;$
($,t$) denote the multiple covariant differentiation according to
the Levi-Civita connection ($t$). We can expand $\cal D$ in a Taylor
series expansion in $t$ to write $\cal D$ invariantly in the form

\bea {\cal D}u=D u+\sum_{r>0} t^r {\cal G}_r^{ij} u_{;ij} +{\cal
F}_r^i u_{;i}  \label{d}\eea

\noindent with the operator $D$ depending on a connection ${\cal
A}_i$ and a smooth function $Q$  given by (\ref{d1}) and the tensor
${\cal G}^{ij}_r$ and ${\cal F}^i_r$ up to the second-order by

\bea
&&{\cal G}_1^{ij}=g^{ij}_{,t}(0)={\rho_{ij,t}(0) \over 2} \sigma_i(0) \sigma_j(0)+\rho_{ij}(0) \sigma_{i,t}(0) \sigma_j(0) \label{g1}\\
&&{\cal G}_2^{ij}={1 \over 2} g^{ij}_{,t t}(0)={\rho_{ij,t t} (0)
\over 4} \sigma_i(0) \sigma_j(0)
+{\rho_{ij,t }(0) \over 2} \sigma_{i,t}(0) \sigma_j(0) \\&&+{\rho_{ij}(0) \over 4} \sigma_{i,t}(0) \sigma_{j,t}(0)   \\
&&{\cal F}_1^i=b^i_{,t }(0) \\
&&{\cal F}_2^i=b^i_{,t t}(0) \eea

\noindent Using this connection, (\ref{fp1}) can be written in the
covariant way, i.e.

\bea {\partial  \over \partial t}p(x,y,t)={\cal D} p(x,y,t)
\label{hk}\eea

\noindent with ${\cal D}$ given by (\ref{d}).

\noindent The asymptotic resolution of the heat kernel (\ref{hk}) in
the short time in a time-dependent background is an important
problem in quantum cosmology.  When the spacetime slowly varies, the
time-dependent metric describing the cosmological evolution can be
expanded in a Taylor series with respect to t. The index $r$ in this
situation is  related to the adiabatic order \cite{dav}. The
following expression obtained in \cite{gil1,gil2}  gives the
complete asymptotic solution for the heat kernel on a Riemannian
manifold.

\begin{thm}
Let M be a Riemannian manifold without a boundary and a
time-dependent metric. Then for each $x \in M$, there is a complete
asymptotic expansion

\bea p(x,y,t)={\sqrt{g(x)} \over (4\pi t)^{n \over
2}}\sqrt{\Delta(x,y)}{\cal P}(x,y)e^{-{\sigma(x,y) \over 2t}}
\sum_{n=0}^\infty a_n(x,y,t)t^n \label{timehk}\eea
\end{thm}

\noindent The $a_i(x,y,t)$ are smooth functions on $M$ and depend on
geometric invariants such as the scalar curvature $R$. The
coefficients ${a}_n$ have been computed up to the fourth-order
($a_0(x,y,t)=1$). The first coefficients are given below

\beaa
&&a_1(x,x,t)=P \equiv {1 \over 6} R+Q +{3 \over 4} {\cal G}_{1,ii} \\
&& a_2(x,x,t)={1 \over 180}(R_{ijkl} R^{ijkl}-R_{ij}R^{ij}+\Delta
R^2)+{1 \over 2} P^2+{1 \over 12} {\cal R}_{ij} {\cal R}^{ij}+{1
\over 6} \Delta P \\&+&{1 \over 360}({45 \over 4} {\cal G}_{1,ii}
{\cal G}_{1,jj} +{45 \over 2}{\cal G}_{1,ij} {\cal G}_{1,ij}+60
{\cal G}_{2,ii}+15{\cal G}_{1,ii} R-30{\cal
G}_{1,ij}R_{ij}\\&&+60{\cal F}_{1,i;i}+15{\cal G}_{1,ii;jj}-30{\cal
G}_{1,ij;ij})  \eeaa

\section{Geometry of Complex Curves and Asymptotic Volatility}

\noindent In our geometric framework, a SVM corresponds to a complex
curve, also called Riemann surfaces. Using the classification of
conformal metric on a Riemann surface, we will show that the SVM
falls into three classes. In particular, the $\lambda$-SABR and
Heston models correspond to the Poincar\'e hyperbolic plane. This
connection between the SABR model and $\HH^2$ has already been
presented in \cite{ber1,sab1,len,len1}. This identification allows
us to find an exact solution to the SABR model ($\lambda=0$) with
$\beta=0,1$. The $\beta=0$ solution  has already been obtained in an
unpublished paper \cite{len,sab1} and rederived by the author in
\cite{phl}. Furthermore, we will derive a general asymptotic implied
volatility for any stochastic volatility models. This expression
only depends on the geometric objects that we have introduced in
this section (i.e. metric, connection).

\subsection{Complex curves}

 \noindent
On a Riemann surface we can show that the metric can always be
written locally in a neighborhood of a point (using the right system
of coordinates)

\bea g_{ij}=e^{\phi(x,y)} \delta_{ij} \;, \; i,j=1,2\eea

\noindent and it is therefore locally conformally flat. The
coordinates $x_i$ are called the isothermal coordinates.
Furthermore, two metrics on a Riemann surface, $g_{ij}$ and $h_{ij}$
(in local coordinates), are called conformally equivalent if there
exists a function $\phi(x_i)$ such that \bea g_{ij}=e^{\phi}
h_{ij}\eea

\noindent The following theorem follows from the above observations:

\begin{thm}[Uniformisation]
Every metric on a simply connected Riemann surface\footnote{The
non-simply connected Riemann surfaces can also be classified by
taking the double cover.} is conformally equivalent to a metric of
constant scalar curvature $R$:
\begin{enumerate}
    \item $R=+1$: the Riemann sphere $S^2$
    \item $R=0$: the complex plane $\mathbb C$
    \item $R=-1$: the upper half plane ${\mathbb H}^2=\{z \in {\mathbb C} | Im(z)>0
    \}$
\end{enumerate}

\end{thm}
\noindent By the uniformisation theorem, all surfaces falls into
these three types and we conclude that  there are a priori three
types of stochastic volatility models (modulo the conformal
equivalence). In the following, we compute the metric associated
with the $\lambda$-SABR model and find the corresponding metric on
$\HH^2$ \cite{len1}. In this way, the $\lambda$-SABR  model can be
viewed as an {\sl universal stochastic volatility model}.
Furthermore, we show that the Heston and $\lambda$-SABR models
belong to the same class of conformal equivalence.

\noindent In the next section, we  present our general asymptotic
implied volatility at the first-order and postpone the derivation to
subsection 4.3.

\subsection{Unified Asymptotic Implied Volatility}

The general asymptotic implied volatility at the first-order (for
any (time-independent) stochastic volatility models) , depending
implicitly on the metric $g_{ij}$ (\ref{metric}) and the connection
${\cal A}_i$ (\ref{aaa}) on our Riemann surface, is given by

\bea &&\sigma_{BS}(K,\tau, g_{ij} ,{\cal A}_i)={ln({K \over f_0})
\over \int_{f_0}^K {df' \over \sqrt{2 g^{ff}(c)}}}(1+{g^{ff} \tau
\over 12}(-{3 \over 4}({\partial_f g^{ff}(c) \over
g^{ff}(c)})^2+{\partial^2_f g^{ff}(c) \over g^{ff}(c)}+{1 \over
f^2}\\&&+{6 \over g^{ff}(c) \phi''(c)}( g^{ff'}(c)( (ln(\Delta g
{\cal P}^2)'(c)-{\phi'''(c) \over \phi''(c)})+g^{ff''}(c)))))
\label{generalimplied}\eea

\noindent with $c$ the volatility $a$ which minimizes the geodesic
distance $d(x,x')$ on the Riemann surface ($\phi(x,x')=d^2(x,x')$).
$\Delta$ is the VanVleck-Morette determinant (\ref{mor}), $g$ is the
determinant of the metric and ${\cal P}$ is the parallel gauge
transport (\ref{pargauge}). \noindent Here the prime symbol $'$
indicates a derivative according to $a$.

\noindent This formula (\ref{generalimplied}) is particularly useful
as we can use it to calibrate rapidly any SVM. In the section 5, we
will apply it to the $\lambda$-SABR model. In order to use the
formula, the only computation needed is the calculation of the
geodesic distance. For example, for a $n$-dimensional hyperbolic
space $\HH^n$, the geodesic distance is known. We will see that
$\HH^2$ corresponds to a SABR model with drift. The general case
$\HH^n$ will be used in a future paper \cite{phl2} to compute an
asymptotic volatility formula for a SABR-BGM model.

\subsection{Derivation}

\subsubsection{Asymptotic probability}

\noindent We now have the necessary data  to apply the heat kernel
expansion and deduce the asymptotic formula for the probability
density at the first-order  for any (time-independent) stochastic
volatility model. We obtain \bea p={1 \over (4\pi
\tau)}\sqrt{\Delta(x,y)}{\cal P}(x,y)e^{-{d^2(x,y) \over 4
\tau}}(1+(Q+{R \over 6}) \tau) \label{asym}\eea

\noindent We will now derive an asymptotic expression for the
implied volatility. The computation involves two steps. The first
step as illustrated in this section consists in computing the local
volatility $\sigma(f,\tau)$ associated to the stochastic SABR model.
In the second step (see next section),  we will deduce the implied
volatility from the local volatility using the heat kernel on a
time-dependent real line.

\noindent We know that the local volatility associated to a SVM is
given by

\bea \sigma^2(f,\tau)=  {2 \int_{0}^\infty \sqrt{g} g^{ff} p da
\over \int_{0}^\infty p \sqrt{g} da} \label{mean} \eea

\noindent with $p$ the conditional probability given in the short
time-limit at the first-order by (\ref{asym}).

\noindent We set $\phi(z,z')=d^2(z,z')$. The factor $\sqrt{g}$ is
the (invariant) measure (see Appendix A) and $g^{ff}={2 \over
\sigma_f^2}$ with the definition (\ref{metric}).

\noindent Plugging our asymptotic expression for the conditional
probability (\ref{asym}) in (\ref{mean}), we obtain

\bea \sigma^2(f,\tau)=  { \int_{0}^\infty f(a,\tau) e^{x \phi(a)}
{da} \over \int_{0}^\infty h(a,\tau) e^{x \phi(a)} {da}}
\label{mean2}\eea

\noindent with $\phi(a)=d^2(z,z')$, $h(a,\tau)=\sqrt{g}
\sqrt{\Delta(z,z')}{\cal P}(z,z')(1+(Q+{R \over 6})\tau)$,
$f(a,\tau)=h(a,\tau) g^{ff}$ and $x=-{1 \over 4 \tau}$.

\noindent Using a saddle-point method, we can find an asymptotic
expression for the local volatility. For example, at the zero-order,
$\sigma^2$ is given by $g^{ff}(c)$ with  $c$ the stochastic
volatility which minimises the geodesic distance on  our Riemann
surface:

\bea c \equiv a \, | \, min_a \phi \label{local0}\eea

\noindent Using the saddle-point method at the first-order, we find
the following expression for the numerator in (\ref{mean2}) (see
Appendix for a sketch of the proof)

\beaa \int_{0}^\infty f(a) e^{x \phi(a)} {da}=\sqrt{{ 2 \pi \over -x
\phi''(c)}} f(c) (1+{1 \over x}( -{ f''(c) \over 2 f(c) \phi''(c)}+{
\phi^{(4)}(c) \over 8 \phi''(c)^2}+{f'(c) \phi'''(c) \over 2
\phi''(c)^2 f(c)}-{5 (\phi'''(c))^2 \over 24 (\phi''(c))^3})) \eeaa

\noindent Computing the denominator in (\ref{mean2}) in a similar
way, we obtain a first-order correction of the local volatility

\beaa \sigma(f,\tau)^2=2 g^{ff}(c) (1+{1 \over x}( -{ 1 \over 2
\phi''(c)}({ f''(c) \over f(c)}-{ h''(c) \over h(c)})+{ \phi'''(c)
\over 2 \phi''(c)^2 } ({ f'(c) \over f(c)}-{ h'(c) \over
h(c)})))\eeaa

\noindent Plugging the expression for $f$ and $g$, we finally obtain

\bea \sigma(f,\tau)=\sqrt{2 g^{ff}(c)} (1+{ \tau \over  \phi''(c)}(
{g^{ff'}(c) \over g^{ff}(c)} ( (ln(\Delta g {\cal
P}^2)'(c)-{\phi'''(c) \over \phi''(c)})+{g^{ff''}(c) \over
g^{ff}(c)} ))) \label{generallocal}\eea

\noindent Here the prime symbol $'$ indicates a derivative according
to $a$. This expression depends only on the metric and the
connection ${\cal A}$ on our Riemann surface.

\noindent The final step is to obtain a relation between a local
volatility function $\sigma(f,t)$ and the implied smile. We will
show in the next section how to obtain such a relation using the
heat kernel expansion on a time-dependent one-dimensional real line
(\ref{timehk}).

\subsubsection{Local Volatility Model and Implied Volatility}

Let's assume we  have a  local volatility model

\bea df=C(f,t) dW_t \; f(t=0)=f_0\eea

\noindent The fair value of a European call option (with maturity
date $\tau$ and strike $K$) is  given by (using an integration by
parts and assuming that $\tau$ is small)

\bea {\cal C}(K,\tau,f_0)=( f_0-K)^++{C^2(K,t=0) \over
2}\int_0^{\tau} dT   p(K,T|f_0)  \label{one} \eea

\noindent with $p(K,T|f_0)$ the conditional probability.

\noindent In our framework, this model corresponds to a
(one-dimensional) real curve endowed with the time-dependent metric
$g_{ff}={2 \over C(f,t)^2}$. For $t=0$ and  for the new coordinate
$u=\sqrt{2}\int{df' \over C(f')}$ (with $C(f,0) \equiv C(f)$), the
metric is flat: $g_{uu}=1$. The distance is then given by the
classical Euclidean distance \bea d(u,u')=|u-u'| \eea

\noindent Furthermore, the connection $\cal A$ (\ref{aaa}) and the
function $Q$ (\ref{qqq}) are given by

\bea &&{\cal A}_f=-{1 \over 2} \partial_f ln(C(f)) \label{ceva1}\\
&&Q={C^2(f) \over 4}[({C^{''} \over C})-{1 \over 2}({C' \over C})^2]
\label{cevqq1}\eea

\noindent The parallel transport is then given by ${\cal
P}(f,f_0)=\sqrt{{C(f_0) \over C(f)}}$

\noindent Furthermore, ${\cal G}$ is given by (\ref{g1})

\bea {\cal G}(K,T)=2
\partial_t ln(C(K,t))|_{t=0} \eea

\noindent The first-order conditional probability (using the heat
kernel expansion on a time-dependent manifold (\ref{timehk}) is then

\beaa p(K,T|f_0,t)={1 \over C(K) \sqrt{2 \pi T}}\sqrt{{C(f_0) \over
C(K)}} e^{{-\sigma(f_0,K)^2 \over 2T} }(1+(Q(K)+{3{\cal G}(K) \over
4})T) \eeaa

\noindent Plugging this expression in (\ref{one}), the integration
over $t$ can be performed  and we obtain

\begin{prop}

\bea {\cal C}(K,\tau,f_0)=( f_0-K)^++ {\sqrt{C(K) C(f_0)\tau} \over
2 \sqrt{2 \pi }}(H_1(\omega) +(Q(K)+{3 {\cal G}(K) \over 4})\tau
H_2(\omega)) \label{optionlocal} \eea

\bea H_1(\omega)&=& 2( e^{-\omega^2} + \sqrt{ \pi \omega^2} ( N(\sqrt{2} |\omega|) -1)) \\
H_2(\omega)&=&{2 \over 3}(e^{-\omega^2}(1-2\omega^2)-{2 |\omega|^{3}
\sqrt{\pi}}( N(\sqrt{2} |\omega|) -1)) \\
\omega&=&\int_{f_0}^K {df' \over \sqrt{2 \tau} C(f')} \eea

\end{prop}

\noindent In the case of a constant volatility, the above formula
reduces to

\begin{exa}[Black-Scholes Vanilla option]
\bea {\cal C}(K,\tau,f_0)=( f_0-K)^++ {\sqrt{K f_0 \sigma_0^2 \tau}
\over 2 \sqrt{2 \pi }}(H_1(\bar{\omega}) +(Q(K)+{3{\cal G}(K) \over
4})\tau H_2(\bar{\omega})) \label{optionlocal1} \eea

\noindent with $\bar{\omega}={ ln({f \over K}) \over \sqrt{2 \tau}
\sigma_{0} }$.
\end{exa}

\noindent By identifying the formula (\ref{optionlocal}) with the
same formula obtained with an implied volatility
$\sigma_0=\sigma(K,T)$ (\ref{optionlocal1}), we deduce

\bea \sigma_{BS}(K,T)={\sqrt{C(K) C(f_0)} \over \sqrt{K
f_0}}{H_1(\omega) \over H_1(\bar{\omega})}(1 +(Q(K)+{3{\cal G}(K)
\over 4})\tau{H_2(\omega) \over H_1(\omega)})+{\sigma_{BS}^2(K,T) T
\over 8}{H_2(\bar{\omega}) \over H_1(\bar{\omega})} \label{rec}\eea

\noindent At the zero-order, we obtain $\omega=\bar{\omega}$ i.e.

\bea \sigma_{BS}^0(K,T)= { ln({K \over f_0}) \over \int_{f_0}^K {df'
 \over C(f')}} \label{ber0} \eea

\noindent The formula (\ref{ber0}) has already been found in
\cite{ber} and we will call it the {\it BBF} relation in the
following. Then using the recurrence equation (\ref{rec}), we
obtain at the first-order

 \bea &&\sigma_{BS}(K,T)= { ln({K \over f_0}) \over \int_{f_0}^K {df'
 \over C(f')}}(1 +{ T \over 3}({1 \over 8}({ ln({K \over f_0}) \over \int_{f_0}^K {df'
 \over C(f')}})^2+Q(K)+{3 {\cal
 G}(K) \over 4}) \\
&& \simeq { ln({K \over f_0}) \over \int_{f_0}^K {df' \over
C(f')}}(1 +{C^2(f) T \over 24} ( 2{ C''(f) \over C(f)}-({C'(f) \over
C(f)})^2+{1 \over f^2}+ 12 {\partial_t C \over C^3(f)} )|_{f={f_0+K
\over 2}}) \label{ber1}
 \eea

\noindent In the case $C(f,t)=C(f)$, we reproduce the asymptotic
implied volatility obtained by Hagan-Woodward (\cite{hag}). Now
plugging the local volatility (\ref{generallocal}) into the implied
volatility (\ref{ber1}), we find (\ref{generalimplied}) and this
achieves our derivation of an asymptotic implied volatility at the
first-order.

\section{$\lambda$-SABR  model and hyperbolic geometry}

\subsection{$\lambda$-SABR model}

The volatility $a$ is not a tradable asset. Therefore, in the
risk-neutral measure, $a$ can have a drift. A popular choice is to
make the volatility process mean-reverting. Therefore, we introduce
the $\lambda$-SABR model defined by the following SDE \cite{ber1}

\beaa df&=&a C(f) dW_1 \label{sabr}\\
da&=&\lambda (a-\bar{\lambda})dt +\nu a dW_2 \\
C(f)&=&f^\beta \;,  \; a(0)=\alpha \; ,  f(0)=f_0\eeaa

\noindent where $W_1$ and $W_2$ are two Brownian processes with
correlation $\rho \; \in \; ]-1,1[$. The stochastic Black volatility
is $\sigma_t=a f^{\beta-1}$. In the following section, we present
our asymptotic smile for the $\lambda$-SABR model and postpone the
derivation to the next section.

\subsection{Asymptotic smile for the $\lambda$-SABR}

\noindent The asymptotic smile (with strike $f$ , maturity date
$\tau$ and spot $f_0$) at the first-order associated to the
stochastic $\lambda$-SABR model is

\bea \sigma_{BS}(f_0,f,\tau)={ ln({f_0 \over f}) \over
vol(q)}(1+\sigma_1({f+f_0 \over 2})\tau) \label{lambdasabr}\eea

\noindent with

\beaa \sigma_1(f)&=&{(a_{min} C(f))^2   \over 24 }( {1 \over
f^2}+{2\partial_{ff}(C(f) a_{min}) \over C(f)a_{min}}
-({\partial_f(C(f)a_{min}) \over C(f) a_{min}})^2)\\&&+{ \alpha
\nu^2  T ln({\cal P})'(a_{min})(1-\rho^2)
\sqrt{cosh(d(a_{min}))^2-1} \over 2  d(a_{min})
 }   \eeaa

\noindent with $q={f^{(1-\beta)}-f_0^{(1-\beta)} \over (1-\beta)}$
($\beta \neq 1$), $vol(q)=
    {1 \over \nu}\log ({-{q}\,\nu  - \alpha \,\rho  +
      {\sqrt{{\alpha }^2 + {{q}}^2\,{\nu }^2 +
          2\,{q}\,\alpha \,\nu \,\rho }}\over \alpha(1-\rho)})$ and $a_{min}(q)={\sqrt{{\alpha}^2 +2\,\alpha\,\nu\,\rho\,q +
{\nu}^2\,q^2}}$.

\noindent Moreover we have

\bea &&ln({{\cal P} \over {\cal P}^{SABR}} )'(a_{min})={\lambda
\over \nu^2}(
G_0(\theta_2(a_{min}),A_0(a_{min}),B)\theta_2'(a_{min})
-G_0(\theta_1(a_{min}),A_0(a_{min}),B)\theta_1'(a_{min})\\&&+A_0'(a_{min})(G_1(\theta_2(a_{min}),B)-G_1(\theta_1(a_{min}),B)))
\label{res4} \eea

\noindent with

\bea &&ln({\cal P}^{SABR})'(a_{min})={\beta \over 2(1-\rho^2)
(1-\beta)} ( F_0(\theta_2(a_{min}),A(a_{min}),B)\theta_2'(a_{min})
\\&&-F_0(\theta_1(a_{min}),A(a_{min}),B)\theta_1'(a_{min})-A'(a)(F_1(\theta_2(a_{min}),A(a_{min}),B)-F_1(\theta_1(a_{min}),A(a_{min}),B)))
\eea

\noindent and with

\beaa &&G_1(x,b)=-\csc (x) + b\,{Re}(\log (\tan (\frac{x}{2})))\\
&&G_0(x,a,b)=\cot (x)\,\csc (x)\,\left( {a} + \sin (x)
\right) \,\left( 1 + b\,\tan (x) \right) \\
&&A_0(a_{min})=-\left( \frac{\bar{\lambda}\,{\sqrt{1 -{\rho}^2}}}{c(f)} \right) \\
&&A_0'(a_{min})=\frac{\bar{\lambda}\,{\sqrt{1 - {\rho }^2}}\,\left(
\alpha \,\rho  + \nu \,q(f) \right) } {{c(f)}^2\,\left( \rho
\,\left( \alpha  - c(f) \right)  + \nu \,q(f) \right)
  } \\
   &&F_0(x,a,b)= \frac{\sin (x)}{a + \cos (x) + b\,\sin (x)}\\
&&F_1(x,a,b)=\frac{-2\,b\,\arctan (\frac{b + \left( -1 + a \right)
\,\tan (\frac{x}{2})}{{\sqrt{-1 + a^2 - b^2}}})}
   {{\left( -1 + a^2 - b^2 \right) }^{\frac{3}{2}}} +
  \frac{-1 + a^2 + a\,b\,\sin (x)}{\left( -1 + a^2 - b^2 \right) \,\left( a + \cos (x) + b\,\sin (x) \right)
  }\\
&&\theta_2(a_{min})= \pi -\arctan (\frac{{\sqrt{1 - {\rho
}^2}}}{\rho
})\\
&&\theta_1(a_{min})=-\arctan (\frac{\alpha \,{\sqrt{1 - {\rho
}^2}}}{\alpha \,\rho  + \nu \,q(f)}) +\pi 1_{(\alpha \,\rho  + \nu \,q(f)) \geq 0}\\
&&\theta_2'(a_{min})={ \alpha \theta_1'(a_{min}) \over a_{min}} \\
&&\theta_1'(a_{min})=\frac{\alpha \,\left( \nu \,q(f) + \rho
\,\left( \alpha  +
         {a_{min}} \right)  \right) }{\nu \,
    {\sqrt{1 - {\rho }^2}}\,q(f)\,\left( 2\,\alpha \,\rho  + \nu \,q(f) \right) \,
    {a_{min}}}\\
&&A(a_{min})=-\left( \frac{\nu \,\left( {f_0} - {{f_0}}^{\beta
}\,\left( -1 + \beta  \right) \,q(f) \right) }
    {{{f_0}}^{\beta }\,\left( -1 + \beta  \right) \,
      {a_{min}}}
      \right) \\
&& A'(a_{min})=\frac{{f_0}\,\nu \,\left( \alpha \,\rho  + \nu \,q(f)
\right)  +
    {{f_0}}^{\beta }\,\alpha \,\left( -1 + \beta  \right) \,\left( \alpha  + \nu \,\rho \,q(f) \right) }{
    {{f_0}}^{\beta }\,\left( -1 + \beta  \right) \,
    a_{min}^2 \,
    \left( \nu \,q(f) + \rho \,\left( \alpha  - {a_{min}} \right)  \right)
    } \\
&& B={\rho \over \sqrt{1-\rho^2}} \eeaa

\subsection{Derivation}

\noindent In order to use our general formula for the implied
volatility, we will compute the metric and the connection associated
to the $\lambda$-SABR model in the next subsection. We will show
that the $\lambda$-SABR metric is diffeomorphic equivalent to the
metric on $\HH^2$, the hyperbolic Poincar\'e plane \cite{len,len1}.

\subsubsection{Hyperbolic Poincar\'e plane}

\noindent The  metric associated to the $\lambda$-SABR model is
(using (\ref{metric}))

\bea ds^2&=&g_{ij}dx^i dx^j \\&=& {2 \over a^2 C^2 \nu^2
(1-\rho^2)}[ \nu^2 df^2-2\nu\rho C(f)dadf+C(f)^2da^2] \eea

\noindent Let's introduce the variable $q(f)=\int_{f_0}^f{df' \over
C(f')}$. By introducing the new coordinates $x=\nu q-\rho a$ and
$y=(1-\rho^2)^{1 \over 2} a$, the metric becomes (after some
algebraic manipulations)  the standard hyperbolic metric on the
Poincar\'e half-plane $\HH^2$ in the coordinates $[x,y]$
\footnote{How can we prove that this is the correct metric on
$\HH^2$? By applying a Moebius transformation (see above), the upper
half plane is mapped to the Poincar\'e disk ${\cal D}=\{ z \in \CC\;
| \; \; |z| \leq 1 \}$. Then if we define $x_1={1+|z|^2 \over
1-|z|^2}$, $x_1={2Re(z) \over 1-|z|^2}$, $x_3={2Re(z) \over
1-|z|^2}$, we obtain that $\cal D$ is mapped to the Minskowski
pseudo-sphere $-x_0^2+x_1^2+x_2^2=1$. On this space, we have the
metric $ds^2=-dx_0^2+dx_1^2+dx_3^2$. We can then deduce the induced
metric on the Minskowki model. On the upper-half plane, this gives
(\ref{meh}) (without the scale factor ${2 \over \nu^2}$).}

\bea  ds^2={2 \over \nu^2}[{dx^2+dy^2 \over y^2}] \label{meh} \eea

\noindent The unusual factor ${2 \over \nu^2}$ in front of the
metric (\ref{meh}) can be eliminated by scaling the time
$\tau'={\nu^2 \over 2}\tau$ in the heat kernel (\ref{hkk}) (and $Q$
becomes ${2 \over \nu^2}Q$). This is what we will use in the
following.

\begin{rem}[Heston model]

\noindent The Heston model is a stochastic volatility model given by
the following SDEs \cite{hes}:

\bea df&=&a f dW_1 \\
da&=&-({\eta^2 \over 8a}+{\lambda a \over 2}(1-({\bar{a} \over
a})^2)dt+{\eta \over 2}dW_2 \eea

\noindent Let's introduce the variable $x={\eta \over 2}ln(f)-\rho
{a^2 \over 2}$, $y=(1-\rho^2)^{1 \over 2} {a^2 \over 2}$. Then, in
the coordinates $[x,y]$, the metric becomes

\bea ds^2={4 \over \eta^2 (1-\rho^2)^{1 \over 2}} y {ds^2_{\HH^2}}
\eea

\noindent and therefore belongs to the same class of conformal
equivalence as $\HH^2$. Modulo this equivalence, the $\lambda$-SABR
and Heston models correspond to the same geometry.

\end{rem}

\noindent As this connection between the $\lambda$-SABR model and
$\HH^2$ is quite intriguing, we investigate some of the useful
properties of the hyperbolic space (for example the geodesics).
First, by introducing the complex variable $z=x+iy$, the metric
becomes

\bea ds^2={ dz d{\bar z} \over Im(z)^2} \eea

\noindent In this coordinate system, it can be shown that
$PSL(2,\RR)$ \footnote{$PSL(2,\RR)=SL(2,\RR)/\ZZ_2$ with $SL(2,\RR)$
the group of two by two real matrices with determinant one. $\ZZ_2$
acts on $A \; \in \; SL(2,\RR)$ by $\ZZ_2 A=-A$} is an isometry,
meaning that the distance is preserved. The action of $PSL(2,\RR)$
on $z$ is transitive and given by

\bea z'={a z +b \over c z +d} \eea

\begin{center}
\begin{figure}[tpb]
\centering

\includegraphics[width=13cm, height=8cm]{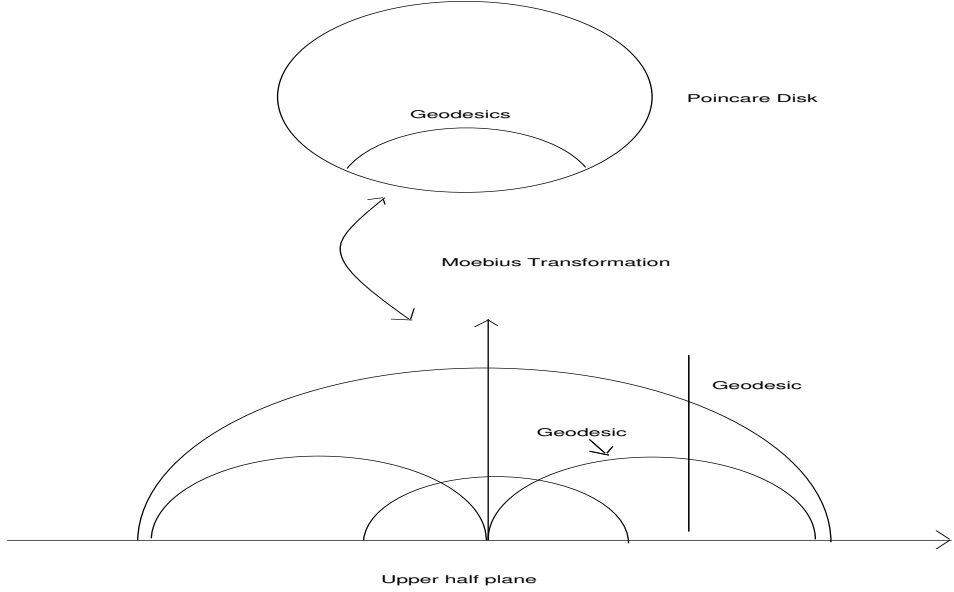}
\caption{Poincar\'e disk $\cal D$ and upper half-plane $\mathbb H$
with some geodesics. In the upper half-plane, the geodesics
correspond to vertical lines and to semi circles centered on the
horizon $Im(z)=0$ and in $\cal D$ the geodesics are circles
orthogonal to $\cal D$} \label{fig2}
 \end{figure}
 \end{center}

\noindent Furthermore, let's define the Moebius transformation $T
\label{mt}$ as an element of $PSL(2,\RR)$ which is uniquely given by
its values at $3$ points: $T(0)=1$, $T(i)=0$ and $T(\infty)=-1$. If
$Im(z)>0$ then $|T(z)|<1$ so $T$ maps the upper half-plane on the
Poincar\'e disk ${\cal D}=\{ z \in \CC\; | \; \; |z| \leq 1 \}$. In
the upper half-plane, the geodesics correspond to vertical lines and
to semi circles centered on the horizon $Im(z)=0$, and in $\cal D$
the geodesics are circles orthogonal to $\cal D$ (Fig. \ref{fig2}).

\noindent Solving the Euler-Lagrange equation (\ref{geodesics}), it
can be proven that the (hyperbolic) distance (invariant under
$PSL(2,\RR)$) between the points $z=x+iy$, $z'=x'+iy'$ on ${\mathbb
H}^2$ is given by (see Appendix A)

\bea d(z,z')=cosh^{-1} [1+{|z-z'|^2 \over 2yy'}] \label{met}\eea

\noindent Using this explicit expression for the hyperbolic
distance, the VanVleck-Morette determinant is

 \bea
\Delta(z,z')={ d(z,z') \over \sqrt{cosh^2(d(z,z'))-1} }
\label{morette}\eea

\noindent Using our specific expression for the geodesic distance
(\ref{met}), we find that $a_{min}$, the saddle-point, is the
following expression

\bea a_{min}(f)={\sqrt{{\alpha}^2 +2\,\alpha\,\nu\,\rho\,q +
{\nu}^2\,q^2}} \eea

 \noindent As $\Delta$ depends only on
the geodesic distance $d$ and $d$ is minimised for $a=a_{min}$, we
have $ln(\Delta)'(a_{min})=0$.

\noindent Furthermore, we have (using Mathematica)

\bea &&a_{min} \phi''(a_{min})={2 d(a_{min}) \over \alpha
 (1-\rho^2)\sqrt{cosh(d(a_{min}))^2-1} } \label{res1}\\
&&{\phi'''( a_{min}) \over \phi'''(a_{min})}=-{3 \over a_{min}} \label{res2}\\
&&d(a_{min})=cosh^{-1}( {-q \nu \rho-\alpha \rho^2+a_{min}(q) \over
\alpha (1-\rho^2)})  \label{res3}\eea

\noindent This above formula (\ref{res3}) has already been obtained
in \cite{ber1}.

\subsubsection{Connection for the $\lambda$-SABR}

\noindent In the coordinates $[a,f]$, the connection ${\cal A}$ is

 \beaa &&{\cal A}={1 \over 2(1-\rho^2)}({( 2 \lambda \bar{\lambda}
 \rho - 2 \lambda
\rho a-\nu a^2 C'(f))  \over \nu  C(f) a^2} df + { (-2 \lambda
\bar{\lambda} +2 \lambda a +\nu \rho a^2 C'(f)) \over \nu^2
 a^2} da) \label{aze} \eeaa

\noindent In the new coordinates $[x,y]$, the above connection is
given by

\bea {\cal A}={\cal A}^{SABR}+{\lambda
(\bar{\lambda}\sqrt{1-\rho^2}-y) \over \nu^2 y^2
\sqrt{1-\rho^2}}(\rho dx-\sqrt{1-\rho^2} dy) \eea

\noindent with  \bea {\cal A}^{SABR}&=&-{\beta \over 2(1-\rho^2)
(1-\beta)} {dx \over
(x+{\rho \over \sqrt{1-\rho^2}} y+{ \nu f_0^{1-\beta} \over (1-\beta)})} \; \; {\mathrm for} \; \; \beta \neq 1 \\
{\cal A}^{SABR}&=&-{dx \over 2(1-\rho^2) \nu } \; \; {\mathrm for}
\; \; \beta =1 \eea

\noindent The pullback of the connection on a geodesic ${\cal C}$
satisfying $(x-x_0(a,f))^2+y^2=R^2(a,f)$ is given by ($\beta \neq 1$
\footnote{The case $\beta=1$ will be treated in the next section})

\bea i^*{\cal A}=i^*{\cal A}^{SABR}+{\lambda
(-\bar{\lambda}\sqrt{1-\rho^2}+y) \over \nu^2 y^2}({\rho y \over
\sqrt{(R^2(a,f)-y^2)(1-\rho^2)} } +1)dy \eea

\noindent and with  \bea i^*{\cal A}^{SABR}={\beta \over 2(1-\rho^2)
(1-\beta)} { y dy \over \sqrt{R^2(a,f)-y^2}
(\hat{x_0}(a,f)+\sqrt{R^2(a,f)-y^2}+{\rho \over \sqrt{1-\rho^2}} y)}
\label{ASABR} \eea

\noindent with $i: {\cal C} \rightarrow \HH^2$ the embedding of the
geodesic ${\cal C}$ on the Poincar\'e plane and $\hat{x}_0=x_0+{ \nu
f_0^{1-\beta} \over (1-\beta)}$. We have used that $i^*dx= -{y dy
\over \sqrt{R^2-y^2}}$.

\noindent Note that the two constants $x_0$ and $R$ are determined
by using the fact that the two points $z_1=-\rho \alpha +i
\sqrt{1-\rho^2} \alpha$ and $z_2=\nu \int_{f_0}^f{df' \over C(f')}
-\rho a +i \sqrt{1-\rho^2}  a$ pass through the geodesic curves. The
algebraic equations given $R$ and $x_0$ can be exactly solved:

\bea &&x_0(a,f)={x_1^2-x^2_2+y_1^2-y_2^2 \over 2(x_1-x_2)} \\
&&R(a,f)={1 \over 2}\sqrt{ {
((x_1-x_2)^2+y_1^2)^2+2((x_1-x_2)^2-y_1^2)y_2^2+y_2^4 \over
(x_1-x_2)^2}} \eea

\noindent Using polar coordinates $x-x_0=R cos(\theta)$, $y=R
sin(\theta)$, we obtain that the parallel gauge transport is

\bea ln({\cal P})(a)=ln({\cal P}^{SABR})(a)+{\lambda \over \nu^2}
\int_{\theta_1(a)}^{\theta_2(a)} (1+B
tan(\theta))(sin(x)+A_0){cos(x) \over sin(x)^2} dx \eea

\noindent with  \bea {\cal
P}^{SABR}(z,z')&=&exp(\int_{\theta_1}^{\theta_2}{\beta \over
2(1-\rho^2) (1-\beta)} {sin(\theta) d \theta \over (cos(\theta)
+{\hat{x}_0 \over R}+{\rho \over \sqrt{1-\rho^2}} sin(\theta))})
\label{para}\eea

 \noindent with
$\theta_i(a,f)=arctan({y_i \over x_i-x_0}) \;,\; i=1,2$,
$A(a,f)={\hat{x}_0(a,f) \over R(a,f)}$, $B={\rho \over
\sqrt{1-\rho^2}}$ and $A_0=-{\bar{\lambda} \sqrt{1-\rho^2} \over
R}$. The two integration bounds $\theta_1$ and $\theta_2$ explicitly
depend on $a$ and  the coefficient ${\hat{x}_0 \over R}$. Doing the
integration, we obtain (\ref{res4}).

\noindent Plugging all these results
(\ref{res1},\ref{res2},\ref{res3},\ref{res4}) in
(\ref{generalimplied}), we obtain our final expression for the
asymptotic smile at the first-order associated to the stochastic
$\lambda$-SABR model (\ref{lambdasabr}).

\begin{rem}[SABR original formula]

We can now see how the classical Hagan-al asymptotic smile
\cite{sab} formula can be obtained in the case $\lambda=0$ and show
that our formula gives a better approximation.

\noindent First we approximate $a_{min}$ by the following expression
\bea a_{min} \simeq \alpha +q \rho \nu  \eea

\noindent In the same way, we have  \bea
{\sqrt{cosh(d(a_{min}))^2-1} \over d(a_{min})} \simeq 1 \eea

\noindent Furthermore, for $\lambda =0$, the connection (\ref{aze})
reduces to

\bea &&{\cal A}={1 \over 2(1-\rho^2)} (-d ln(C(f)) +{\rho \over \nu}
\partial_f C da) \label{saba}\eea

\noindent Therefore, the parallel gauge transport is obtained by
integrating this one-form along a geodesic $\cal C$

 \bea  {\cal P}=exp( {1 \over 2(1-\rho^2)}( -ln({C(f) \over
 C(f_0)})+\int_{\cal C} {\rho \over \nu} \partial_f C da))
 \eea

\noindent The component $f$ of the connection is an exact form and
therefore has easily been integrated. The result doesn't depend on
the geodesic but only on the endpoints. However, this is not the
case for the component ${\cal A}_a$. But by approximating
$\partial_f' C(f') \simeq \partial_f C(f)$ , the component ${\cal
A}_a$ becomes an exact form and can therefore be integrated

 \bea \int_{\cal C} {\rho \over \nu}
\partial_f C da \simeq {\rho \over \nu} \partial_f C(f) (a-\alpha)
\eea

\noindent Finally, plugging these approximations into our formula
(\ref{generalimplied}), we reproduce the Hagan-al original formula
\cite{sab}

\bea \sigma_{BS}(f_0,f,\tau)={ ln({f_0 \over f}) \over
vol(q)}(1+\sigma_1({f+f_0 \over 2})\tau) \eea

\noindent with  \beaa \sigma_1(f)&=&{(\alpha C(f))^2   \over 24 }(
{1 \over f^2}+{2\partial_{ff}(C(f) ) \over C(f)} -({\partial_f(C(f))
\over C(f) })^2)+{\alpha \nu  \partial_f(C(f))\rho \over 4}+{2-3
\rho^2 \over 24} \nu^2  \eeaa

\noindent Therefore, the Hagan-al classical formula corresponds to
the approximation of  the Abelian connection by an exact form. The
latter can be integrated  outside the parametrisation of the
geodesic curves.

\end{rem}

\begin{figure}
\centering
\includegraphics[width=13cm, height=8cm]{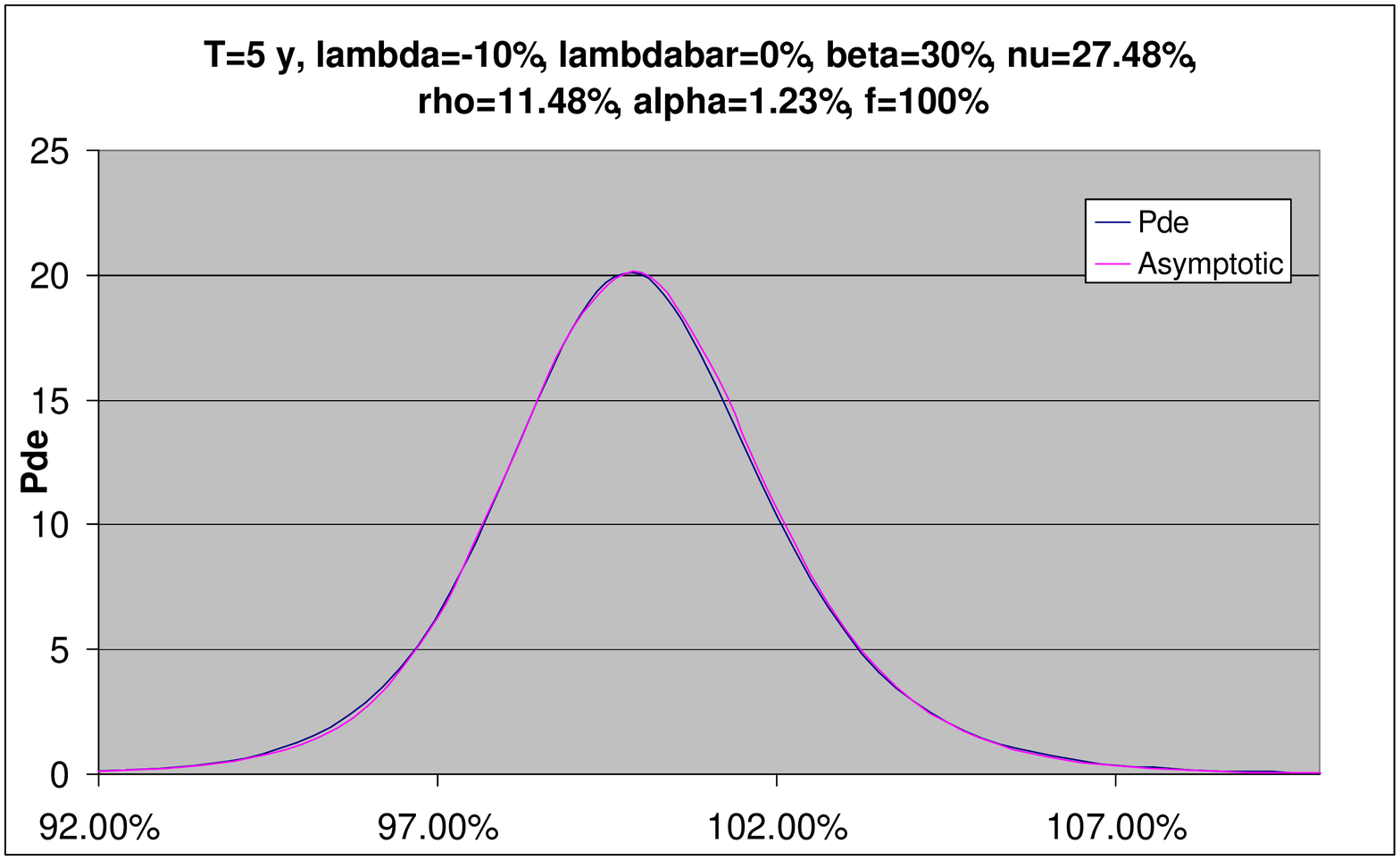}
\includegraphics[width=13cm, height=8cm]{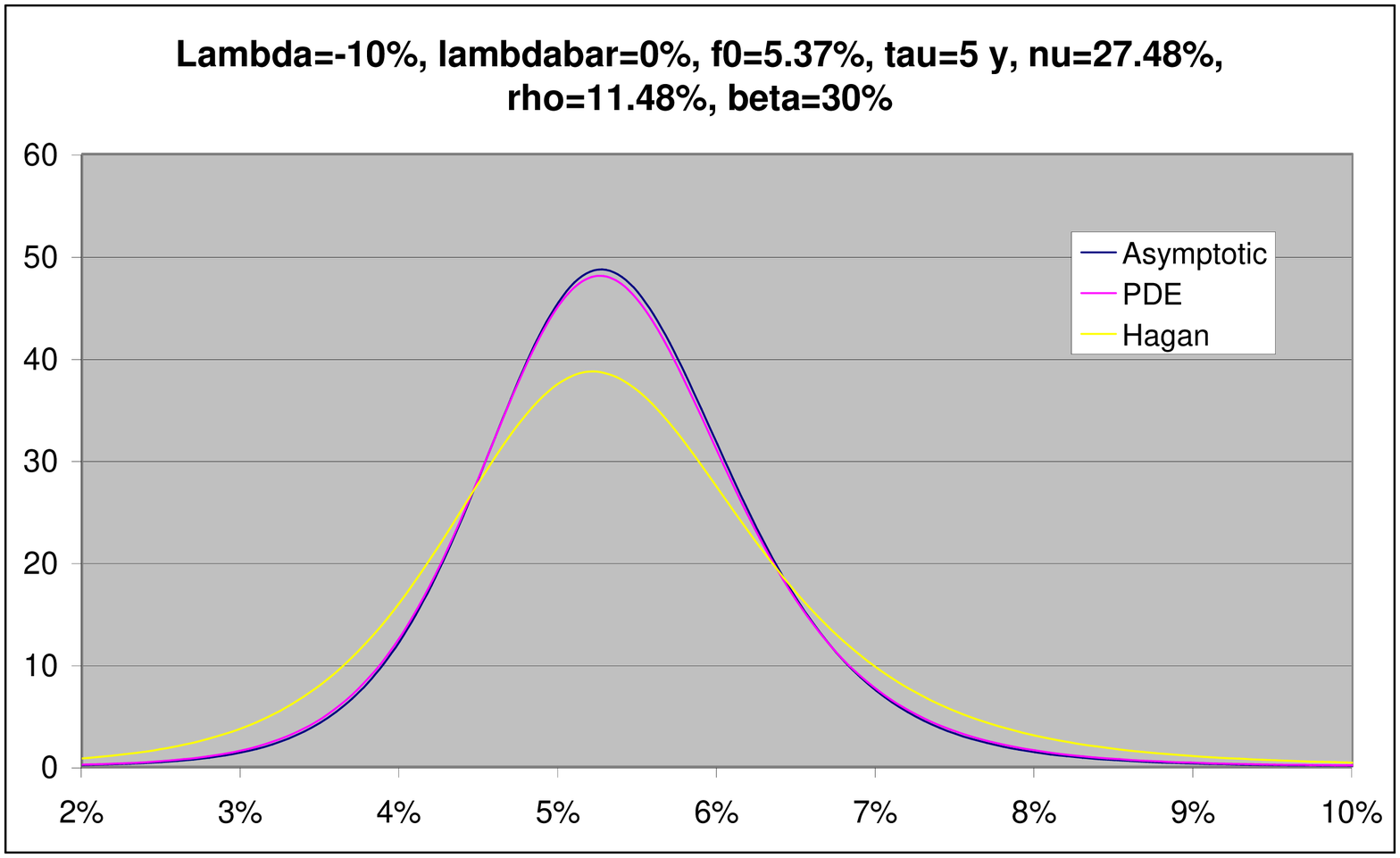}

\end{figure}

\begin{figure}
\centering
\includegraphics[width=13cm, height=8cm]{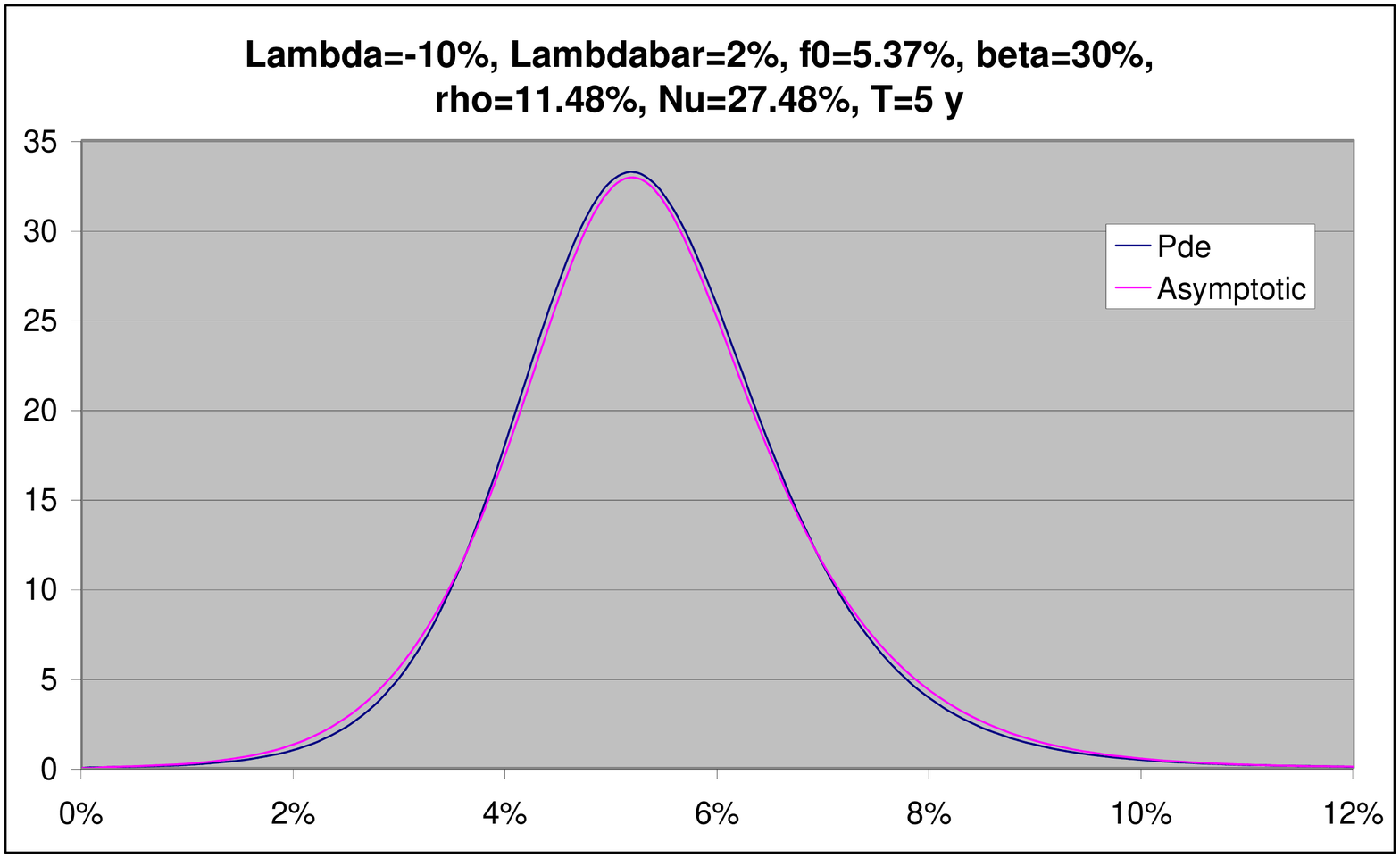}
\caption{Pdf $p(K,T|f_0)={\partial^2 {\cal C} \over \partial^2 K}$.
Asymptotic solution (pink) vs numerical solution (blue) (pde
solver). In some cases, the Hagan original formula has been plotted
to see the impact of the mean-reversion term.} \label{fig3}
\end{figure}

\begin{rem}[$\HH^2$-model]

\noindent In the previous section, we have seen that the
$\lambda$-SABR mode corresponds to the geometry of $\HH^2$. This
space is particularly nice in the sense that the geodesic distance
and the geodesic curves are known. A similar result holds if we
assume that $C(f)$ is a general function ($C(f)=f^\beta$ for
$\lambda$-SABR). In the following, we will try to fix this arbitrary
function in order to fit the {\it short-term smile}. In this case,
we can use our unified asymptotic smile formula at the zero-order.
The short-term smile will be automatically calibrated by
construction if

\bea \sigma_{loc}(f)=C(f) a_{min}(f) \label{equ1}\\
a_{min}(f)=\alpha^2+2 \rho \alpha \nu q +\nu^2 q^2 \\
q=\int_{f_0}^f {df' \over C(f')} \eea

\noindent with $\sigma_{loc}(f)$ the short-term local volatility. By
{\it short-term}, we mean a maturity date less than $5$ years.
Solving (\ref{equ1}) according to $q$, we obtain

 \bea \nu q= -\rho \alpha \nu +\sqrt{
\alpha^2(-1+\rho^2)+{\sigma_{loc}(f)^2 \over C(f)^2}} \eea

\noindent and if we derive under $f$, we have (
$\psi(f)={\sigma_{loc}(f) \over C(f)}$)

\bea {d \psi \over \sqrt{ \psi^2-\alpha^2(1-\rho^2)}}={\nu \over
\sigma_{loc}(f)}df \eea

\noindent Solving this ODE, we obtain that $C(f)$ is fixed to

 \bea
C(f)={\sigma_{loc}(f) \over \alpha \sqrt{1-\rho^2} cosh[ \nu
\int_{f_0}^f {df' \over \sigma_{loc}(f')}]} \eea

\noindent Using the BBF formula, we have

 \bea C(f)={f
\sigma_{BS}(f) (1-f ln({f \over f_0}){ \sigma_{BS}'(f) \over
\sigma_{BS}(f)})
 \over \alpha \sqrt{1-\rho^2} cosh[ \nu { ln({f \over f_0}) \over
\sigma_{BS}(f)}]} \eea

\noindent Using this function for the $\lambda$-SABR model, the
short term smile is then automatically calibrated.

\end{rem}

\section{Analytical solution for the SABR model with $\beta=0,1$}

\subsection{SABR model with $\beta=0$ and $\HH^ 2$}

\vskip 2truemm

\noindent For the SABR model, the connection ${\cal A}$ and the
function $Q$ are given by

 \bea &&{\cal A}={1 \over 2(1-\rho^2)}
(-\partial_f ln(C) df +{\rho \over \nu} \partial_f C da) \label{newa}\\
&&Q={a^2 \over 4}({ C \partial_f^2 C } -{(\partial_f C)^2 \over
2(1-\rho^2)}) \label{newq}\eea

\noindent For $\beta=0$, the function $Q$  and the potential $\cal
A$ vanish. Then $p$ satisfies a heat kernel where the differential
operator $D$ reduces to the Laplacian on $\HH^2$:

\bea {\partial p \over \partial \tau'}&=&\Delta_{\HH^2} p \label{sel}\\
&=&y^2(\partial_x^2+\partial_y^2)p \eea

\noindent Therefore solving the Kolmogorov equation for the SABR
model with $\beta=0$ (called SAR0 model) is equivalent to solving
this (Laplacian) heat kernel on $\HH^2$. Surprisingly, there is an
analytical solution for the heat kernel on $\HH^2$ (\ref{sel}) found
by McKean \cite{kea}. It is connected to the Selberg trace formula
\cite{gut}.

\noindent The {\it exact conditional probability} density $p$
depends on the hyperbolic distance $d(z,z')$ and is given by (with
$\tau'={\nu^2 \tau \over 2}$)

\beaa p(d,\tau')=2^{-{5 \over 2}} \pi^{-{3 \over 2}} \tau'^{-{3
\over 2}} e^{-{\tau' \over 4}} \int_{d(z,z')}^\infty { b e^{-{b^2
\over 4\tau'}} \over (cosh b -cosh d(z,z'))^{1 \over 2}} db
\label{cp} \eeaa

\noindent The conditional probability in the old coordinates $[a,f]$
is

\beaa &&p(f,a,\tau')dfda={\nu \over (1-\rho^2)^{1 \over 2}} {df da
\over a^2 C(f)} 2^{-{5 \over 2}} \pi^{-{3 \over 2}} \tau'^{-{3 \over
2}} e^{-{\tau' \over 4}} \int_{d(z,z')}^\infty { b e^{-{b^2 \over
4\tau'}} \over (cosh b -cosh d(z,z'))^{1 \over 2}} db \eeaa

\begin{figure}
\includegraphics[width=13cm, height=8cm]{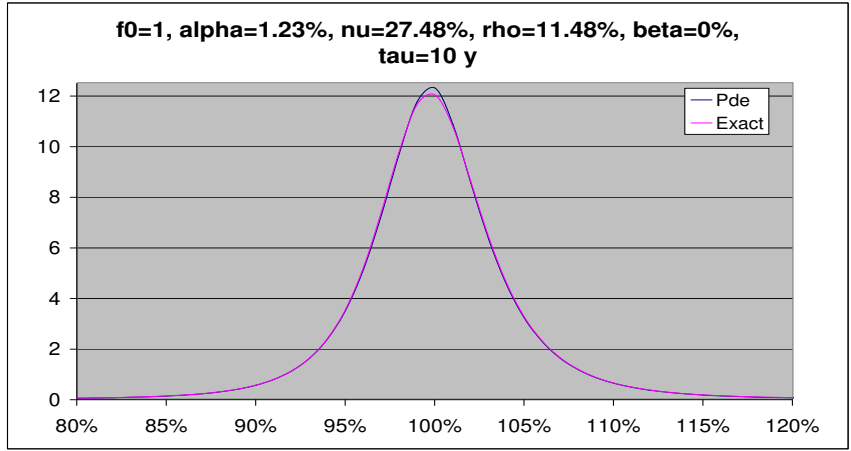}
\caption{Conditional probability for the SABR model with $\beta=0$
versus numerical PDE.} \label{sar0p}
 \end{figure}

\noindent We have compared this exact solution (Fig. \ref{sar0p})
with a numerical PDE solution of the SAR0 model and found agreement.
A similar result was obtained previously in \cite{len,sab1} for the
conditional probability.

 \noindent The value of a European option is  (after an integration
by parts)

\beaa C(f,K)=(f-K)^++{1 \over 2} \int_0^{\tau_{ex}} d\tau
\int_{0}^{\infty}
  da  a^2 p(x_1=K,a,\tau|\alpha) \eeaa

\noindent In order to integrate over $a$ we use a small trick: we
interchange the order of integration over $b$ and $a$. The half
space $b \geq d$ with $a$ arbitrary is then mapped to the half-strip
$a_{min} \leq a \leq a_{max}$ and $b \geq l_{min}$ where
\footnote{all these algebraic computations have been done with
Mathematica}

\beaa &&a_{max}-a_{min}=2{\sqrt{\,{( \alpha\,( ch({b}) +{\rho}^2
-ch({b})\,{\rho}^2 )  - \nu\,\rho\,q ) }^2 -
     \,( {\alpha}^2 - 2\,\alpha\,\nu\,\rho\,q + {\nu}^2\,q^2 )
     }} \\
&&a_{min}+a_{max}= 2 \alpha ch(b) + 2 \alpha \rho^2 (1-ch(b)) + 2
\nu \rho q
\\ &&cosh({l_{min}})= {num \over den}
     \eeaa

\noindent with

\beaa &&num=-{\alpha}^2 - 2\,{\alpha}^2\,{\rho}^2 -
2\,{\alpha}^2\,{\rho}^4 +
  4\,{\alpha}^2\,{\rho}^6 - 6\,\alpha\,\nu\,{\rho}^3\,s +
  4\,\alpha\,\nu\,{\rho}^5\,s - 3\,{\nu}^2\,{\rho}^2\,s^2 \\&&+
  2\,{\nu}^2\,{\rho}^4\,s^2 +
  \rho\,\left( -1 + 2\,{\rho}^2 \right) \,
   \left( \alpha\,\rho + \nu\,s \right) \,
   {\sqrt{{\alpha}^2 + 2\,\alpha\,\nu\,\rho\,
        \left( -1 + 2\,{\rho}^2 \right) \,s +
       {\nu}^2\,\left( -1 + 2\,{\rho}^2 \right) \,s^2}} \\
       &&den=\alpha\,\left( -1 + {\rho}^2 \right) \,\left( 1 +
2\,{\rho}^2 \right) \,
  \left( 2\,\rho\,\left( \alpha\,\rho + \nu\,s \right)  +
    {\sqrt{{\alpha}^2 + 2\,\alpha\,\nu\,\rho\,
         \left( -1 + 2\,{\rho}^2 \right) \,s +
        {\nu}^2\,\left( -1 + 2\,{\rho}^2 \right) \,s^2}} \right)
\eeaa

\noindent Performing the integration according to $a$ leads to an
{\it exact solution} for the SABR model with $\beta=0$

\beaa C(f,K)=(f-K)^++{1 \over \nu (1-\rho^2)^{1 \over 2}}2^{-{5
\over 2}} \pi^{-{3 \over 2}}\int_0^{{\nu^2 \tau_{ex} \over 2}} dt
  t^{-{3 \over
2}} e^{-{t \over 4}} \int_{l_{min}}^\infty { b (
a_{max}-a_{min})e^{-{b^2 \over 4t}} \over \sqrt{(cosh b -cosh
l_{min})}} db \eeaa

\subsection{SABR  model with $\beta=1$ and the three-dimensional hyperbolic space $\HH^3$}
A similar computation can be carried out for $\beta=1$.  Using
(\ref{newa}), we can show that the potential $\cal A$ is exact,
meaning there exists a smooth function $\Lambda$ such that ${\cal
A}=d\Lambda$ with $\Lambda={1 \over 2(1-\rho^2)}(-ln(f)+{\rho \over
\nu}a)$. Furthermore, using (\ref{newq}), we have $Q=-{a^2 \over
8(1-\rho^2)}=-{y^2  \over 8(1-\rho^2)^2}$.

\noindent Applying an Abelian gauge transformation $p'=e^{\Lambda}p$
(\ref{gt}), we find that $p'$ satisfies the following equation

\bea y^2(\partial_x^2+\partial_y^2 -{1 \over 4
\nu^2(1-\rho^2)^2})p'=\partial_{\tau'} p' \label{h3}\eea

\noindent How do we solve this equation? It turns out that the
solution corresponds in some  fancy way to the solution of the
(Laplacian) heat kernel on the three dimensional hyperbolic space
$\HH^3$. This space can be represented as the upper-half space
$\HH^3=\{ x=(x_1,x_2,x_3) | x_3>0 \}$. In these coordinates, the
metric takes the following form

\bea ds^2={(dx_1^2+dx_2^2+dx_3^2) \over x_3^2} \eea

\noindent and the geodesic distance between two points $x$ and $x'$
in $\HH^3$ is given by \footnote{$|\cdot|$ is the Euclidean distance
in $\RR^3$}

\bea cosh(d(x,x'))=1+{|x-x'|^2 \over 2x_3x_3'} \eea

\noindent As in $\HH^2$, the geodesics are straight vertical lines
or semi-circles orthogonal to the boundary of the upper-half space.
An interesting property, useful to solve the heat kernel, is that
the group of isometries of $\HH^3$ is $PSL(2,\CC)$
\footnote{$PSL(2,\CC)$ is identical to $PSL(2,\RR)$, except that the
real field is replaced by the complex field.}. If we represent a
point $p \in \HH^3$ as a quaternion \footnote{The quaternionic field
is generated by the unit element $\bf 1$ and the basis $\bf i$, $\bf
j$, $\bf k$ which satisfy the multiplication table ${\bf i}.{\bf
j}={\bf k}$ and the other cyclic products.}  whose fourth components
equal zero, then the action of an element $g \in PSL(2,\CC)$ on
$\HH^3$ can be described by the formula \bea p'=g.p={ ap+b \over
cp+d} \eea \noindent with $p=x_1{\bf 1}+x_2{\bf i}+x_3{\bf j}$.

\noindent The Laplacian on $\HH^3$ in the coordinates
$[x_1,x_2,x_3]$ is given by

\bea
\Delta_{\HH^3}=x_3^2(\partial_{x_1}^2+\partial_{x_2}^2+\partial_{x_3}^2)
\eea

\noindent and the (Laplacian) heat kernel is

\bea \partial_{\tau'} p'=\Delta_{\HH^3}p'\label{cvb}\eea

\noindent The exact solution for the conditional probability density
$p'(d(x,x'),t)$, depending on the geodesic distance $d(x,x')$, is
\cite{gri}

\bea p'(d(x,x'),\tau')={1 \over (4\pi \tau')^{3 \over 2}} {d(x,x')
\over sinh(d(x,x'))} e^{-\tau'-{d(x,x')^2 \over 4\tau'}} \eea

\noindent Let's apply a Fourier transformation on $p$ along the
coordinate $x_1$ (or equivalently $x_2$) \bea
p'(x_1,x_2,x_3,x',\tau')=\int_{-\infty}^\infty {dk \over \sqrt{2
\pi}} e^{ikx_1} \hat{p}(k,x_2,x_3,x',t) \eea

\noindent Then $\hat{p}'$ satisfies the following PDE

\bea
\partial_{\tau'}\hat{p}'=x_3^2(-k^2+\partial_{x_2}^2+\partial_{x_3}^2)\hat{p}'
\label{h3bis} \eea

\noindent By comparing (\ref{h3bis}) with (\ref{h3}), we deduce that
the exact solution for the conditional probability for the SABR
model with $\beta=1$ is (with $x \equiv x_2$, $y \equiv x_3$, $k
\equiv {1 \over 2\nu (1-\rho^2)}$, $x_1' =0$)

\beaa p'(x,y,x',y',\tau')=e^{-{1 \over 2(1-\rho^2)}(ln({f \over
K})-{\rho \over \nu}(a-\alpha))} \int_{-\infty}^\infty {dx_1 \over
\sqrt{2 \pi}}{1 \over (4\pi \tau')^{3 \over 2}} {d(x,x') \over
sh(d(x,x'))} e^{-\tau'-{d(x,x')^2 \over 4 \tau'}} e^{-{ix_1 \over 2
\nu (1-\rho^2)}} \eeaa

\noindent A previous solution for the SABR model with $\beta=1$ was
obtained by \cite{lew}, although only in terms of Gauss
hypergeometric series.

\section{Conclusions and future work}

\noindent Let's summarize our findings. By using the heat kernel
expansion, we have explained how to obtain a general asymptotic
smile formula at the first-order. As an application, we have derived
the smile formula for a SABR model with a mean-reversion term.
Furthermore, we have shown how to  reproduce \cite{hag,sab}. In the
case of the SABR model with $\beta=0, 1$, exact solutions have been
found, corresponding to the geometries of $\HH^2$ and $\HH^3$,
respectively. We note that these solutions are not easy to obtain
without exploiting this connection with hyperbolic geometry.

\vskip 3truemm

\noindent This geometric framework  allows to obtain other
analytical solutions for stochastic volatility models. For example,
a solution for the heat kernel on a multi-dimensional hyperbolic
space $\HH^n$ exists. In a future work, we will show that this space
corresponds to a BGM model coupled with a SABR model. Using the
general heat kernel expansion on a manifold, we will derive an
asymptotic smile for this particular model \cite{phl2}.

\section*{Acknowledgements}
I would like to thank Dr. C. Waite and Dr. G. Huish for stimulating
discussions. I would also like to acknowledge Prof. Avramidi for
bringing numerous references on the heat kernel expansion to my
attention. Moreover, I would like to thank Dr. A. Lesniewski and
Prof. H. Berestycki for bringing their respective papers
\cite{ber1,sab1,len1} to my attention.

\section{Appendix: Notions in differential geometry}

\subsection*{Riemanian manifold}

\includegraphics[width=13cm, height=5cm]{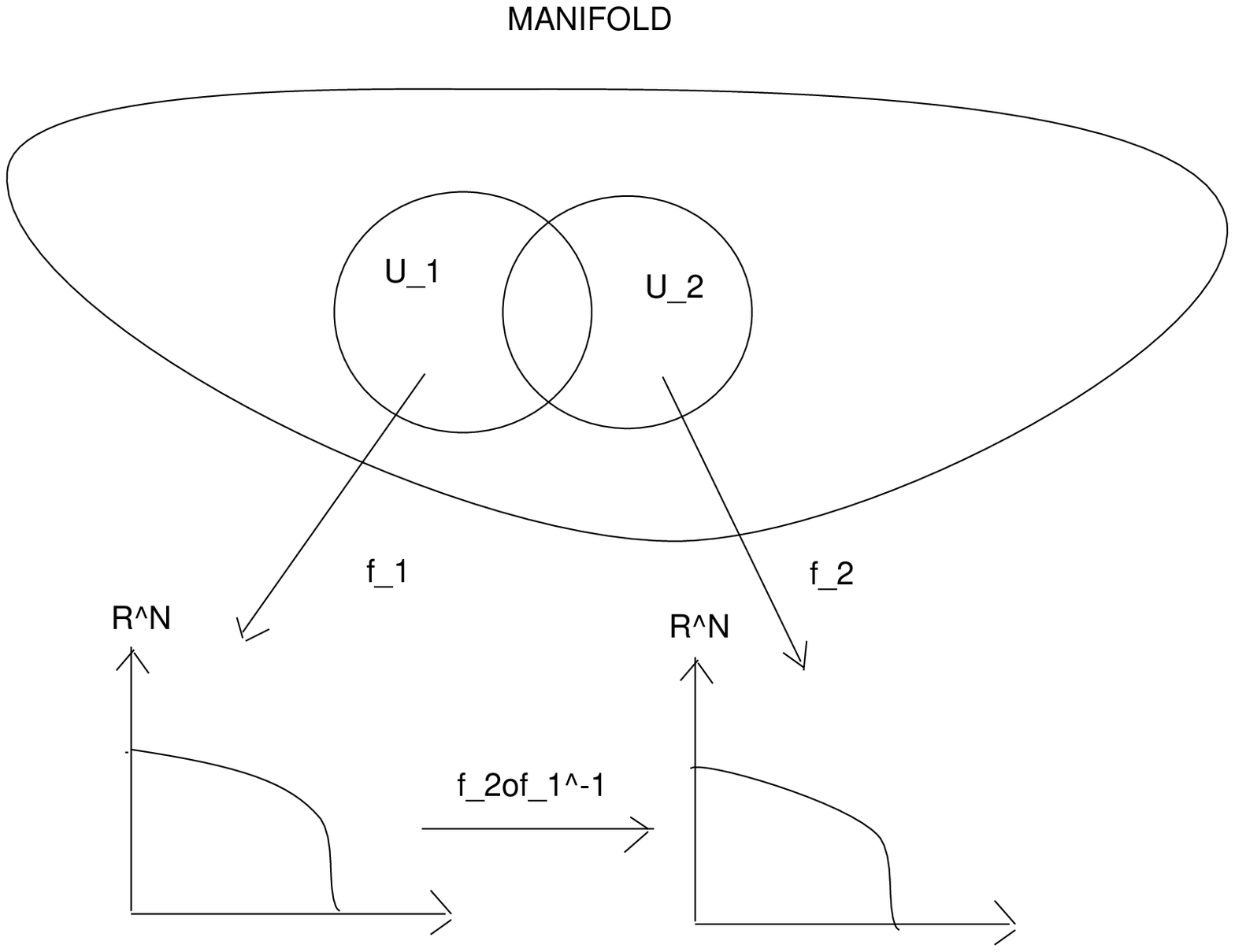}

A real $n$-dimensional manifold is a space which looks like $\RR^n$
around each point. More precisely, $M$ is covered by open sets
${\cal U}_i$ (topological space) which are homeomorphic to $\RR^n$
meaning that there is a continuous map $\phi_i$ (and its inverse)
from $\UU_i$ to $\RR^n$ for each $i$. Furthermore, we impose that
the map $\phi_{i,j}=\phi_i^{-1} o \phi_j$ from $\RR^n$ to $\RR^n$ is
$C^{\infty}(\RR^n)$.

\noindent As an example, a two-sphere $S^2$ can be covered with two
patches:  ${\cal U}_N$ and ${\cal U}_S$, defined respectively as
$S^2$ minus the north pole, and the south pole. We obtain the map
$\phi_N$ ($\phi_S$) by doing a stereographic projection on $\UU_N$
($\UU_S$). This projection  consists in taking the intersection of a
line passing through the North (South) pole and a point $p$ on $S^2$
with the equatorial plane. We can show that $\phi_{SN}(x,y)=({x
\over (x^2+y^2)}, -{y \over (x^2+y^2)})$ is $C^\infty$ and even
holomorphic. So $S^2$ is a {\it complex manifold}.

\includegraphics[width=13cm, height=5cm]{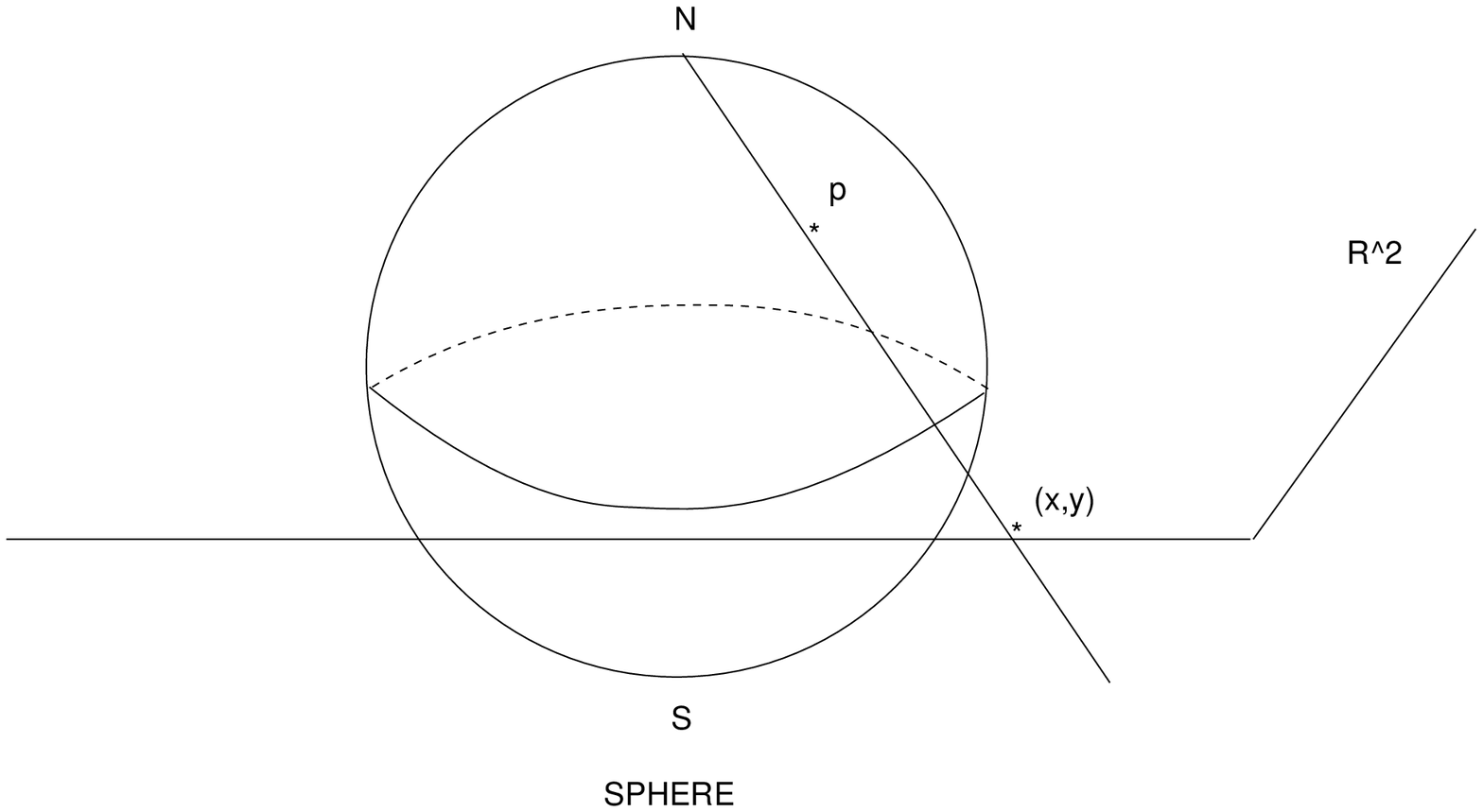}

\subsection*{Metric}

A metric $g_{ij}$ written with the local coordinates $x_i$
(corresponding to a particular chart $\UU_i$) allows us to measure
the distance between infinitesimally nearby points $x^{i}$ and
$x^i+dx^i$: $ds^2=g_{ij}dx^i dx^j$. If a point $p$ belongs to two
charts then the distance can be computed using two different systems
of coordinates $x^i$ and $x^{i'}=f(x^{i})$. However, the result of
the measure should be the same, meaning that
$g_{ij}dx^idx^j=g_{ij}dx^{i'} dx^{j'}$. We deduce that under a
change of coordinates, the metric is not invariant but changes in a
contravariant way by $g_{ij}=g_{i'j'}\partial_i x^{i'}
\partial_j x^{j'}$.

\noindent A manifold endowed with an Euclidean metric is called a
{\it Riemannian manifold}.

\noindent On a $n$-dimensional Riemannian manifold, the measure
$\sqrt{g} \prod_{i=1}^n dx^i$ is invariant under an arbitrary change
of coordinates. Indeed the metric changes as $g_{ij}=g_{i'j'}
{\partial x^{i'} \over \partial x^i} {\partial x^{j'} \over
\partial x^j}$ and therefore $g=det(g_{ij})$ changes as $\sqrt{g}=det({\partial x^{i'} \over \partial
x^i})\sqrt{g'}$. Furthermore, the element $\prod_{i=1}^n dx^i$
changes as $\prod_{i=1}^n dx^i= det^{-1}({\partial x^{i'} \over
\partial x^i}) \prod_{i'=1}^n dx^{i'}$ and we deduce the result.

\subsection*{Line bundle and Abelian connection}

\subsubsection*{Line bundle}

\includegraphics[width=13cm, height=5cm]{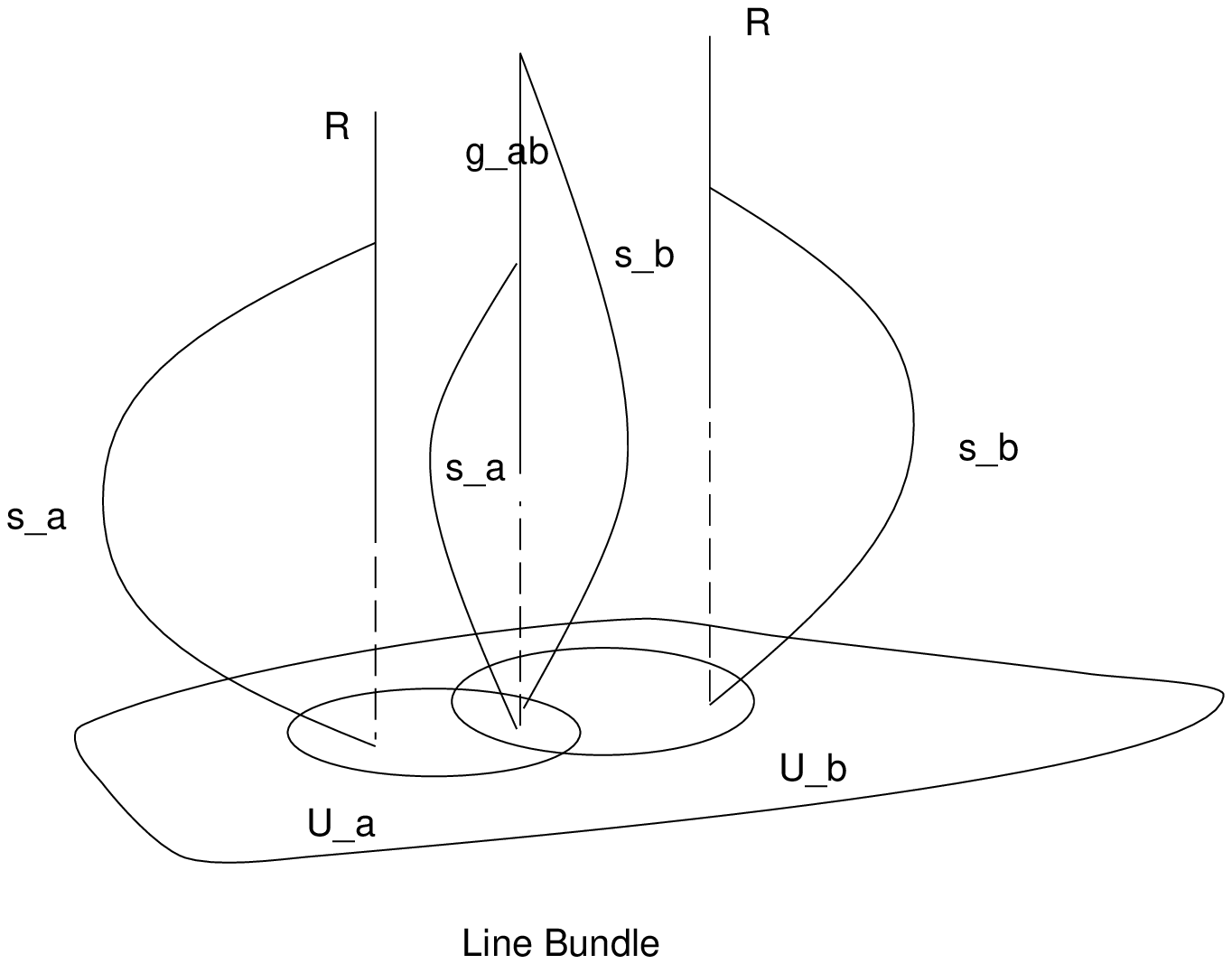}

\noindent A line bundle $\cal L$ is defined by an open covering of
$M$, $\{ {\cal U}_\alpha \}$, and for each $(\alpha,\beta)$, a
smooth transition function $g_{\alpha \beta}: {\cal U}_\alpha \cap
{\cal U}_\beta \rightarrow \RR$ which satisfies the "cocycle
condition" \bea g_{\alpha \beta} g_{\beta \gamma} g_{\gamma
\alpha}=1 \; \; on \; \; {\cal U}_\alpha \cap {\cal U}_\beta \cap
{\cal U}_\gamma \eea

\noindent A section $\sigma$ of ${\cal L}$ is defined by its local
representatives $\sigma$ on each ${\cal U}_\alpha$:

\bea \sigma|_{{\cal U}_\alpha} \equiv \sigma_\alpha : {\cal
U}_\alpha \rightarrow \RR \eea
 \noindent and they are related to each other by the formula
 $\sigma_\alpha=g_{\alpha \beta} \sigma_\beta $ on ${\cal U}_\alpha \cap {\cal U}_\beta$.

\subsubsection*{Abelian connection}
\noindent An Abelian connection $\nabla$ on the line bundle $\cal L$
is a collection of differential operator $\partial_i +{\cal
A}_i^{\alpha}$ on each open set ${\cal U}_\alpha$ which transforms
according to

\bea {\cal A}_i^\alpha={\cal A}_i^\beta +g_{\alpha \beta}
\partial_i g_{\alpha \beta}^{-1} \; \; on \; \; {\cal U}_\alpha
\cap {\cal U}_\beta \eea

\section{Appendix: Laplace integrals in many dimensions}
\noindent Let $\omega$ be a bounded domain on $\RR^n$, $S: \Omega
\rightarrow \RR$, $f: \Omega \rightarrow \RR$ and $\lambda >0$ be a
large positive parameter. The Laplace's method consists in studying
the asymptotic as $\lambda \rightarrow \infty$ of the
multi-dimensional Laplace integrals

\bea F(\lambda)=\int_{\Omega} f(x) e^{\lambda S(x)} dx \eea

\noindent Let $S$ and $f$ be smooth functions and the function $S$
have a maximum only at one interior nondegenerate critial point $x_0
\in \Omega$. Then in the neighborhood at $x_0$ the function $S$ has
the following Taylor expansion

\bea S(x)=S(x_0)+{1 \over 2}(x-x_0)^2 \partial_x^2 S(x_0)
(x-x_0)+o((x-x_0)^3) \eea

\noindent As $\lambda \rightarrow \infty$, the main contribution of
the integral comes from the neighborhood of $x_0$. Replacing the
function $f$ by its value at $x_0$, we obtain a Gaussian integrals
where the integration over $x$ can be performed. One gets the
leading asymptotic of the integral as $\lambda \rightarrow \infty$

\bea F(\lambda) \sim e^{\lambda S(x_0)} ({2 \pi \over \lambda})^{n
\over 2} [-det (\partial^2_x S(x_0))]^{-1 \over 2} f(x_0) \eea

\noindent More generally, doing a Taylor expansion at the
$n^{\mathrm th}$ order for $S$ (resp. $n-2$-order for $f(x)$) around
$x=x_0$, we obtain

\bea F(\lambda) \sim e^{\lambda S(x_0)} ({2 \pi \over \lambda})^{n
\over 2} [-det (\partial^2_x S(x_0))]^{-1 \over 2}
\sum_{k=0}^{\infty} a_k \lambda^{-k} \eea

\noindent with the coefficients $a_k$ are expressed in terms of the
derivatives of the functions $f$ and $S$ at the point $x_0$. For
example, at the first-order (in one dimension), we find

\bea F(\lambda) \sim \sqrt{ {2 \pi \over -\lambda S''(c)}}e^{\lambda
S(x_0)}\{ f(x_0)+{1 \over \lambda}(-{f''(x_0) \over 2 S''(x_0)}+{
f(x_0) S^{(4)}(x_0) \over 8 S''(x_0)^2}+{ f'(x_0) S^{(3)}(x_0) \over
2 S''(x_0)^2}-{5 S'''(x_0) f(x_0) \over 24 S''(x_0)^3}) \}
\label{saddle}\eea

\end{document}